\documentclass[twoside]{article}
\usepackage{amsmath}
\usepackage{amssymb}
\usepackage{graphicx}
\usepackage{amsmath}
\usepackage{amsfonts}
\usepackage{euscript}
\input epsf
\usepackage{amsthm}

\def\enddemo{\qed \endtrivlist}
\expandafter\let\csname enddemo*\endcsname=\enddemo

\def\qedsymbol{\ifmmode\bgroup\else$\bgroup\aftergroup$\fi
  \vcenter{\hrule\hbox{\vrule
height.6em\kern.6em\vrule}\hrule}\egroup}
\def\qed{\ifmmode\else\unskip\nobreak\fi\quad\qedsymbol}

 \newtheorem{thm}{Theorem}[section]
 
 \newtheorem{lem}{Lemma}[section]
 
 \newtheorem{exm}{Example}[section]

 \newtheorem{alg}{Algorithm}[section]

\newcommand{\ra}[1]{\textrm{rank}({#1})}  

\font\ssr=cmss8

\font\sst=cmtt8

\font\ssi=cmti8
 \font\ssr=cmss8  \font\sst=cmtt8
  \font\ssi=cmti8

\title{\bf Symbolic computation of \\weighted Moore-Penrose inverse \\ using partitioning method}
\begin{footnotesize}
\author{\frenchspacing
\bf Milan B. Tasi\' c\footnote{Corresponding author\ }\ , Predrag S. Stanimirovi\' c, Marko D. Petkovi\' c\\
{\ssi University of Ni\v{s}, Department of Mathematics, Faculty of Science,}\\
{\ssi Vi\v segradska 33, 18000 Ni\v s, Serbia} \\
{\ssi E-mail:} {\sst milan12t@ptt.yu},\ \ {\sst pecko@pmf.ni.ac.yu},\ \ {\sst dexter\_of\_nis@neobee.net}\\
}
\end{footnotesize}
\date{}
\begin{document}

\maketitle

\begin{abstract}
We propose a method and algorithm for computing the weighted Moore-Penrose inverse of one-variable rational matrices.
Continuing this idea, we develop an algorithm for computing the weighted Moore-Penrose inverse of
one-variable polynomial matrix.
These methods and algorithms are generalizations of the method or computing the weighted Moore-Penrose inverse
for constant matrices, originated in \cite{Wang}, and
the partitioning method for computing the Moore-Penrose inverse of rational and polynomial matrices introduced in \cite{Stanimirovic2}.
Algorithms are implemented in the symbolic computational package {\ssr MATHEMATICA\/}.

\frenchspacing \itemsep=-1pt
\begin{description}
\item[] AMS Subj. Class.: 15A09, 68Q40.
\item[] Key words: Weighted Moore-Penrose inverse; Rational and polynomial matrices.
\end{description}
\end{abstract}

\section{Introduction}

Let $\mathbf C$ be the set of complex numbers, $\mathbf C^{m\times n}$
be the set of $m \times n $ complex matrices, and
${\mathbf C}^{m\times n}_r\!=\!\{X\in {\mathbf C}^{m \times n}\, :\,\,\, \ra{X}\!=\!r\}$.
For any matrix $A\in\mathbf {C}^{m \times n}$ and positive definite
matrices $M$ and $N$ of the orders $m$ and $n$ respectively, consider the following
equations in $X$, where $*$ denotes conjugate and transpose:
$$\begin{array}{ll}
  (1)\qquad \   AXA=A & (2)\qquad XAX\! =\! X \\
  (3M)\quad  (MAX)^*=MAX & (4N)\quad (NXA)^*\! =\! NXA.
\end{array}$$
The matrix $X$ satisfying (1), (2), (3M) and (4N) is called the weighted Moore-Penrose inverse of  $A$, and it is denoted by $X=A_{MN}^{\dagger}$.
Especially, in the case $M=I_m$ and $N=I_n$, the matrix $X=A_{MN}^{\dagger}$ comes to
the Moore-Penrose inverse of $A$, and it is denoted by $X=A^{\dagger}$.

For any matrix $A\in {\mathbf C}^{n\times n}$ the Drazin inverse of
$A$ is the unique matrix, denoted by $A^D$, and satisfying the
matrix equation (2) and the following equation in $X$:
$$(1^k)\quad \! A^kXA\! =\! A^k,\qquad (5) \quad AX\! =\! XA.$$

As usual, $\mathbf C[s]$ (resp. $\mathbf C(s)$) denotes the polynomials (resp. rational functions) with complex
coefficients in the indeterminate $s$. The $m\times n$ matrices
with elements in $\mathbf C[s]$ (resp. $\mathbf C(s)$) are
denoted by $\mathbf C[s]^{m\times n}$ (resp $\mathbf C(s)^{m\times n}$).
By $I$ is denoted an appropriate identity matrix.

\smallskip
We observed three different directions in the symbolic computation of generalized inverses:

{\bf A)} extensions of Leverrier-Faddeev algorithm,

{\bf B)} methods based on the interpolation, and

{\bf C)} methods based on the Grevile's recursive algorithm.

\smallskip
{\bf A)}
Computation of the Moore-Penrose inverse of one variable polynomial and/or rational matrices, based on the
Leverrier-Faddeev algorithm, is investigated in \cite{Bar,Fra,Jon,Kar,Tze,Pecko1}.
These papers are based on the paper \cite{Decell}.
Implementation of the algorithm from \cite{Kar} in the symbolic computational
language {\ssr MAPLE\/}, is described in \cite{Jon}.
An algorithm for computing the Moore-Penrose inverse of two-variable rational and polynomial matrix is introduced in \cite{Kar4}.
A quicker and less memory-expensive effective algorithm
for computing the Moore-Penrose inverse of one-variable and two-variable polynomial matrix,
with respect to those introduced in \cite{Kar} and \cite{Kar4}, is presented in \cite{Kar2}.
This algorithm is efficient when elements of the input matrix are polynomials with only few nonzero addends.

\smallskip
Continuing the algorithm of the Leverrier-Faddeev type for computing the Drazin inverse of constant matrices, established in \cite{Gr},
a representation and corresponding algorithm
for computing the Drazin inverse of a nonregular polynomial matrix of an arbitrary degree
is introduced in \cite{Ji}, \cite{Pecko, Pecko1}.
Bu and Wei in \cite{Bu} proposed a finite algorithm for symbolic computation of the Drazin
inverse of two-variable rational and polynomial matrices.
Also, a more effective three-dimensional version of these algorithms is presented in the paper \cite{Bu}.
Implementation of this algorithm in the programming language {\ssr MATLAB\/} is presented in \cite{Bu}.

\smallskip
A general for of the Leverrier-Faddev type algorithms is introduced in \cite{Pecko2}.
This algorithm generates the class of outer inverses of a rational or polynomial matrix.

\smallskip
{\bf B)}
An interpolation algorithm for computing the Moore-Penrose inverse of a given one-variable polynomial matrix, based on the
Leverrier-Faddeev method, is presented in \cite{Marko}.
Algorithms for computing the Moore-Penrose and the Drazin inverse of one-variable polynomial matrices based on
the evaluation-interpolation technique and the Fast Fourier transform are introduced in \cite{Kar3}.
Corresponding algorithms for two-variable polynomial matrices are introduced in \cite{Vol}.
These algorithms are efficient when the input matrix is dense.

\smallskip
{\bf C)}
Grevile's {\it partitioning method\/} for numerical computation of generalized inverses
is introduced in \cite{Gre}.
Two different proofs for Greville's method were presented in \cite{Campbell}, \cite{Wang1}.
A simple derivation of the Grevile's result has been given by Udwadia and Kalaba \cite{Udwadia}.
In \cite{Fan} Fan and Kalaba used the approach of determination of the Moore-Penrose inverse of matrices
using dynamic programming and Belman's principle of optimality.
Wang in \cite{Wang} generalizes Grevile's method to the weighted Moore-Penrose inverse.
Also, the results in \cite{Wang} are proved using a new technique.

\smallskip
In \cite{Shi} the Greville's algorithm is estimated as the method which needs more operations
and consequently it accumulates more rounding errors.
Moreover, it is well-known that the Moore-Penrose inverse is not
necessarily a continuous function of the elements of the matrix. The existence of this
discontinuity present further problems in the pseudoinverse computation. It is therefore clear that
cumulative round off errors should be totally eliminated. During the symbolic implementation,
variables are stored in the "exact" form or can be left "unassigned" (without numerical values),
resulting in no loss of accuracy during the calculation \cite{Kar}.

\smallskip
An algorithm for computing the Moore-Penrose inverse of one-variable polynomial and/or rational matrices,
based on the Grevile's partitioning algorithm, was introduced in \cite{Stanimirovic2}.
An extension of results from \cite{Stanimirovic2} to the set of two-variable rational and polynomial
matrices is introduced in the paper \cite{Stanimirovic3}.

\smallskip
In the present paper we extend Wang's partition method from \cite{Wang} to the set of
one-variable rational and polynomial matrices.
In this way, we obtain an algorithm for computing the weighted Moore-Penrose inverse
of one-variable rational and polynomial matrices.
The paper is a generalization of the paper \cite{Wang} and a continuation of the paper \cite{Stanimirovic2}.

\smallskip
The structure of the paper is as follows.
In the second section we extend the algorithm for computing the weighted Moore-Penrose from \cite{Wang}
to the set of one-variable rational matrices.
In Section 3 we give the main theorem and adapt this algorithm to the set of polynomial matrices.
Several symbolic examples are arranged in fourth section. In partial case $M=I_m$, $N=I_n$
we obtain the usual Moore-Penrose inverse, and then use test examples from \cite{Zielke}.
In the last section we describe main implementation details.

\section{Weighted Moore-Penrose inverse for rational matrices} \setcounter{equation}{0}

Greville in \cite{Gre} proposed the partitioning algorithm which relates
the Moore-Penrose pseudoinverse of a constant matrix $R$ augmented by a
vector $r$ of appropriate dimensions with the pseudoinverse $R^\dagger $ of $R$.
Wang and Chen in \cite{Wang} generalize Greville's partitioning method. They obtained
an algorithm for computing the weighted Moore-Penrose inverse, and give a new technique for its proof.
This method is also suitable for the weighted least-squares problem.

\smallskip
By $\widehat{A}_i(s)$ we denote the submatrix of $A(s)\in \mathbf C(s)^{m\times n}$ consisting of its first $i$ columns:
\begin{equation}\label{dva1}
\widehat{A}_i(s)=\left[ \widehat{A}_{i-1}(s)\ |\ a_i(s)\right],i=2,\ldots ,n,\quad \widehat{A}_1(s)=a_1(s)
\end{equation}
\noindent where $a_i(s)$ is the $i$-th column of $A$.

\smallskip
In the sequel we consider positive definite matrices $M(s)\in\mathbf {C}(s)^{m \times m}$ and $N(s)\in\mathbf {C}(s)^{n \times n}$.
The leading principal submatrix $N_i(s)\in\mathbf {C}(s)^{i\times i}$ of $N(s)$ is partitioned as
\begin{equation}\label{dva2}
    N_i(s)=\left[
               \begin{array}{ll}
                 N_{i-1}(s) & l_i(s) \\
                 l_i^*(s) & n_{ii}(s) \\
               \end{array}
             \right] ,\ i=2,\ldots ,n.
\end{equation}
In the following lemma we generalize the representation of the weighted Moore-Penrose inverse from \cite{Wang}
to the set of one-variable rational matrices.

\smallskip
For the sake of simplicity, by $X_i(s)$ we denote the weighted Moore-Penrose inverse
corresponding to submatrices $\widehat{A}_i(s)$ and $N_i(s)$:
$X_i(s)=\widehat{A}_i(s)^\dagger_{MN_i}$, for each $i=1,\ldots ,n$.

\begin{lem}\label{lem1}
Let $A(s)\in\mathbf {C}(s)^{m \times n}$, assume that $M(s)\in\mathbf {C}(s)^{m \times m}$ and $N(s)\in\mathbf {C}(s)^{n \times n}$ are
positive definite matrices,
and let $\widehat{A}_i(s)$ be the submatrix of $A(s)$ consisting of its first $i$ columns.
Assume that the leading principal submatrix of $N(s)$, denoted by
$N_i(s)\in\mathbf {C}(s)^{i \times i}$, is partitioned as in $(\ref{dva2})$.

In the case $i=1$ we have
\begin{equation}\label{dva3}
X_1(s)=\!\!a_{1}(s)^{\dagger} =
\left\{ \begin{array}{lc}
  \left( a_{1}^*(s) M(s) a_{1}(s)\right) ^{-1} a_{1}^*(s) M(s), & a_{1}(s)\ne 0,\\
  a_1^*(s), & a_{1}(s)=0.
\end{array}
\right.
\end{equation}

For each $i=2,\ldots ,n$ $X_i(s)$ is equal to
\begin{equation}\label{dva4}
    X_{i}(s)\!\!=\!\!\left[
          \begin{array}{c}
            X_{i-1}(s)\!-\!\left( d_i(s)+(I-X_{i-1}(s)\widehat{A}_{i-1}(s)\right)   N_{i-1}^{-1}(s)l_i(s))b_i^*(s) \\
            b_i^*(s) \\
          \end{array}
        \right]
\end{equation}
where the vectors $d_i(s)$, $c_i(s)$ and $b_i^*(s)$ are defined by
\begin{eqnarray}
  d_i(s) &=& X_{i-1}(s)a_i(s)\label{dva5} \\
  c_i(s) &=& a_i(s)-\widehat{A}_{i-1}(s)d_i(s) = \left( I-\widehat{A}_{i-1}(s)X_{i-1}(s)\right) a_i(s).\label{dva6}
\end{eqnarray}

\begin{equation}\label{dva7}
    b_i^*(s)=\left\{
       \begin{array}{ll}
             \left( c^*_i(s) M(s) c_i(s)\right) ^{-1}c_i^*(s) M(s), & c_i(s)\ne 0 \\\\
             \delta_i^{-1}(s)\left( d_i^*(s) N_{i-1}(s)-l_i(s)^*\right) X_{i-1}(s), & c_i(s)=0,
       \end{array}
     \right.
\end{equation}
and where in $(\ref{dva7})$ is
\begin{eqnarray}\label{dva8}
    \delta_i(s)&=&n_{ii}(s)+d^*_i(s) N_{i-1}(s) d_i(s)-\left( d^*_i(s) l_i(s)+l_i^*(s)d_i(s)\right) \nonumber\\
    &&\ -l_i^*(s)\left( I-X_{i-1}(s) \widehat{A}_{i-1}(s)\right) N_{i-1}^{-1}(s)l_i(s).
\end{eqnarray}
\end{lem}

\begin{demo}
The proof is the same as for constant matrices, presented in \cite{Wang}.
\end{demo}

The following lemma is a simple extension of the well-known result in the literature.

\begin{lem}\label{lem20}
Let $A(s)$ be a partitioned matrix which is nonsingular, and let the submatrix $A_{11}(s)$ also be nonsingular. Then
\begin{eqnarray}\label{dva9}
A(s)=\left[\begin{array}{ll}
             A_{11}(s) & A_{12}(s) \\
             A_{21}(s) & A_{22}(s)
       \end{array}\right]^{-1}=
\left[
       \begin{array}{ll}
             B_{11}(s) & B_{12}(s) \\
             B_{21}(s) & B_{22}(s)
       \end{array}
     \right]
\end{eqnarray}
where
\begin{eqnarray}
B_{11}(s)&=&A_{11}(s)^{-1}+A_{11}(s)^{-1}A_{12}(s)B_{22}(s)A_{21}(s)A_{11}(s)^{-1}\label{b11}\label{dva10}\\
B_{12}(s)&=&-A_{11}(s)^{-1}A_{12}(s)B_{22}(s)\label{b12}\label{dva11}\\
B_{21}(s)&=&-B_{22}(s)A_{21}(s)A_{11}(s)^{-1}\label{b21}\label{dva12}\\
B_{22}(s)&=&(A_{22}(s)-A_{21}A_{11}(s)^{-1}A_{12}(s))^{-1}. \label{dva13}
\end{eqnarray}
\end{lem}

The following lemma is a generalization of known result from \cite{Wang} to the set of rational matrices.

\begin{lem}\label{lem2}
Let $N_i(s) $ be the partitioned matrix defined in $(\ref{dva2})$. Assume that $N_i(s)$ and $N_{i-1}(s)$ are both nonsingular. Then
\begin{eqnarray} \label{dva14}
N_i^{-1}(s)=\left[\begin{array}{ll}
             N_{i-1}(s) & l_i(s) \\
             l_i^*(s) & n_{ii}(s)
       \end{array}\right]^{-1}=
\left[
       \begin{array}{ll}
             E_{i-1}(s) & f_i(s) \\
             f_i^*(s) & g_{ii}(s)
       \end{array}
     \right]
\end{eqnarray}
where
\begin{eqnarray}
g_{ii}(s)&=&\left( n_{ii}(s)-l_i^*(s)N_{i-1}^{-1}(s)l_i(s)\right) ^{-1} \label{dva15}\\
f_i(s)&=&-g_{ii}(s)N_{i-1}^{-1}(s)l_i(s)\label{dva16}\\
E_{i-1}(s)&=&N_{i-1}^{-1}(s)+g_{ii}^{-1}(s)f_i(s)f_i^*(s).\label{dva17}
\end{eqnarray}
\end{lem}

\begin{demo}
The proof immediately follows from the substitutions $A_{11}(s)=N_{i-1}(s)$, $A_{12}(s)=l_i(s)$, $A_{21}(s)=l_i(s)^*$ and $A_{22}(s)=n_{ii}(s)$
in {\it Lema \ref{lem20}}.
\end{demo}

In view of Lemma {\ref {lem1}} we present the following algorithm for
computing the weighted Moore-Penrose inverse of a given one-variable rational matrix.

\begin{alg}\label{alg21}

Input: rational matrix $A(s)\in \mathbf C(s)^{m\times n}$ and positive definite matrices
$M(s)\in \mathbf C(s)^{m\times m}$ and $N(s)\in \mathbf C(s)^{n\times n}$.

 \smallskip
\leftskip 0.5cm
{\it Step 1.\/} Initial value: Compute $X_1(s)=a_1(s)^{\dagger}$ defined in $(\ref{dva3})$.

\smallskip
{\it Step 2.\/} Recursive step: For each $i=2,\ldots ,n$  compute $X_i(s)$ performing the following four steps:

\leftskip 0.8cm

\smallskip
{\it Step 2.1.\/} Compute $d_i(s)$ using $(\ref{dva5})$.

{\it Step 2.2.\/} Compute $c_i(s)$ using $(\ref{dva6})$.

{\it Step 2.3.\/} Compute $b_i^*(s)$ by means of $(\ref{dva7})$ and $(\ref{dva8})$.

{\it Step 2.4.\/} Applying $(\ref{dva4})$ compute $X_i(s)$.

\leftskip 0.5cm

\smallskip
{\it Step 3.\/} The stopping criterion: $i=n$. Return $X_n(s)$.

\leftskip 0cm

\end{alg}

Let $N_i(s)$, defined in $(\ref{dva2})$,
be the leading principal submatrix of positive definite matrix $N(s)\in \mathbf C(s)^{n\times n}$.
The following algorithm, based on Lemma {\ref {lem2}}, computes the inverse matrix $N_i^{-1}(s)\in \mathbf C(s)^{i\times i}$.

\begin{alg}\label{alg22}
Compute $N_i^{-1}(s)$.

\smallskip
\leftskip 0.5cm
{\it Step 1.\/} Initial values: $N_{1}^{-1}(s)=n_{11}^{-1}(s)$

\leftskip 0.5cm

\smallskip
{\it Step 2.\/} Recursive step: For $i=2,\ldots ,n$ perform the following steps:

\leftskip 0.8cm

{\it Step 2.1.\/} Compute $g_{ii}(s)$ using $(\ref{dva15})$.

{\it Step 2.2.\/} Compute $f_i(s)$ applying $(\ref{dva16})$.

{\it Step 2.3.\/} Compute $E_{i-1}(s)$ according to $(\ref{dva17})$.

{\it Step 2.4.\/} Compute $N_i^{-1}(s)$ using $(\ref{dva14})$.

\leftskip 0.5cm
\smallskip
{\it Step 3.\/} Stopping criterion: for $i=n$ the output is the inverse matrix $N^{-1}(s)=N_n^{-1}(s)$.

\leftskip 0cm

\end{alg}

\section{Weighted Moore-Penrose inverse for polynomial matrices} \setcounter{equation}{0}

Consider the matrix $A(s)\in \mathbf {C}[s]^{m \times n}$ given in the polynomial form with
respect to unknown $s$:
\begin{eqnarray}\label{tri1}
  A(s)=A_1+A_2s+\cdots +A_qs^{q-1}+A_{q+1}s^q=\sum\limits_{i=0}^q A_{i+1}s^i
\end{eqnarray}
where $A_i$, $i=1,\ldots ,q+1$ are constant $m\times n$ matrices.

\begin{thm}\label{tpoly}
Consider an arbitrary polynomial matrix
$A(s)\in \mathbf C[s]^{m\times n}$ given by $(\ref{tri1})$ and the following polynomial forms of positive definite matrices
$M(s)\in \mathbf C[s]^{m\times m}$ and $N(s)\in \mathbf C[s]^{n\times n}$:
\begin{eqnarray}\label{tri2}
  M(s)=\sum\limits_{i=0}^{m_q} M_{i+1}s^i,\quad N(s)=\sum\limits_{i=0}^{n_q} N_{i+1}s^i.
\end{eqnarray}
Transcribe $i$-th column of $A(s)$ by
\begin{eqnarray}\label{tri3}
a_i(s)\!=\!\sum\limits_{j=0}^q a_{i,j+1} s^j,\quad 1\!\leq \!i\!\leq \!n,\end{eqnarray}
where $a_{i,j+1}$, $0\leq j\leq q$ are constant $m\times 1$ vectors.
Also, denote first $i$ columns of $A(s)$ by
\begin{equation}\label{tri4}
\widehat{A}_i(s)\!=\!\sum\limits_{j=0}^q \widehat{A}_{i,j+1}s^j,\quad 1\!\leq \!i\!\leq \!n,
\end{equation}
where $\widehat{A}_{i,j+1}$, $0\leq j\leq q$ are constant $m\times i$ matrices.

\smallskip
In the partition $(\ref{dva2})$
of the leading principal submatrix $N_i(s)\in\mathbf {C}[s]^{i \times i}$ of $N(s)$, we use the following polynomial representations:
\begin{eqnarray}\label{tri5}
  n_{ii}(s)\!=\!\sum\limits_{j=0}^{n_q} \widehat{n}_{i,j+1}s^j, \quad l_i(s)\!= \! \sum\limits_{j=0}^{n_q} L_{i,j+1}s^j,\quad
  N_{i-1}^{-1}(s) \!= \!\frac{\sum\limits_{j=0}^{\overline{n}_q} \overline{N}_{i-1,j+1}s^j}
                      {\sum\limits_{j=0}^{\overline{\overline{n}}_q} \overline{\overline{N}}_{i-1,j+1}s^j}.
\end{eqnarray}

Then the following algorithm computes the weighted Moore-Penrose inverse $A(s)_{M,N}^\dagger$.

\smallskip
\begin{alg}\label{algpoly}$\ $

\smallskip
\noindent {\it Step 1.\/} Initial values:

\smallskip
\noindent Compute $Z_{1,j+1}$, $0\leq j\leq q_1=q+m_q$ and $Y_{1,j+1}$, $0\leq j\leq p_1=2q+m_q$ as in
\begin{eqnarray}\label{tri6}
Z_{1,j+1}\!\!\!&=\!\!\!\! &\left\{ \begin{array}{lc}
                  \!\!\!\sum\limits_{k=0}^{j} a_{1,j-k+1}^*M_{k+1}, \ 0\leq j\leq q+m_q, & a_1(s)\! \neq \!0,\\ \\
                  a_{1,j+1}^*\!=\!0, & a_{1}(s)\!=\!0.
                  \end{array}
\right.\\
Y_{1,j+1}\!\!\!\!\!&=&\!\!\!\left\{\!\! \!\begin{array}{lc}
          \sum\limits_{r=0}^{j} \sum\limits_{k=0}^{j-r} \!a_{1,j-k-r+1}^* M_{k+1}a_{1,r+1}, \ 0\!\leq \!j\!\leq \!2q\!+\!m_q,& a_1(s)\!\neq \!0,
                                                  \label{tri7}\\ \\
          1, & a_{1}(s)\!=\!0.
          \end{array}
\right.
\end{eqnarray}

\noindent {\it Step 2.\/} Recursive step:

\smallskip
For $2\leq i\leq n$ perform {\it Step 2.1\/}, {\it Step 2.2\/}, {\it Step 2.3\/} and {\it Step 2.4\/}.

\leftskip 0.4cm

\smallskip
{\it Step 2.1.\/} Compute $d_{i,j+1}, 0\leq j\leq q+q_{i-1}$  by means of
\begin{equation}\label{tri8}
\begin{array}{ll}
d_{i,j+1} =\sum\limits_{k=0}^j Z_{i-1,j-k+1}a_{i,k+1},\quad
0\leq j \leq q_{i-1}+q.
\end{array}
\end{equation}

\leftskip 0.4cm
{\it Step 2.2.\/} Compute $c_{i,j+1}$, $0\leq j\leq \hat{q}_{i-1}+q$ using

\leftskip 0cm
\begin{equation}\label{tri9}
c_{i,j+1} \!\!=\!\!\sum\limits_{k=0}^j(a_{i,j-k+1}Y_{i-1,k+1}-
        \widehat{A}_{i-1,j-k+1}d_{i,k+1}),\ 0 \!\leq \! j \! \leq \!\!\hat{q}_{i-1}\!+\!q.
\end{equation}
where

\begin{equation}\label{tri10}
\hat{q}_{i-1} = \max\{ p_{i-1},q_{i-1}+q\}.
\end{equation}

\leftskip 0.4cm
{\it Step 2.3.\/} If $c_{i,j+1}\!\neq \!0$ for some $j$, compute $V_{i,j+1}$
and $W_{i,j+1}$ by means of

\leftskip 0cm
\begin{eqnarray}\label{tri11}
V_{i,j+1}\!&=&\!\sum\limits_{r=0}^{j}\sum\limits_{k=0}^{j-r}Y_{i-1,j-k-r+1}c_{i,k+1}^*M_{r+1},\\
&& \quad 0 \!\leq\! j \!\leq\! \overline{b}_i=\hat{q}_{i-1}+q+p_{i-1}+m_q,\nonumber\\
W_{i,j+1}\!&=&\!\sum\limits_{r=0}^{j}\sum\limits_{k=0}^{j-r}c_{i,j-k-r+1}^*M_{k+1}c_{i,r+1}, \label{tri12}\\
 & & \quad 0\! \leq j \leq {\overline{\overline{b}}_i}=2\hat{q}_{i-1}+2q+m_q.\nonumber
\end{eqnarray}

\leftskip 0.4cm
In the case $c_{i,j+1}=0$ for each $j$, compute $V_{i,j+1}$ and $W_{i,j+1}$ in this way:

\leftskip 0cm
\begin{eqnarray}\label{tri13}
V_{i,j+1}\!\!\!\!&=&\!\!\!\!\sum\limits_{t=0}^{j}\sum\limits_{r=0}^{j-t}
                    \sum\limits_{k=0}^{j-t-r}{\overline{\Delta}}_{i,j-k-r-t+1}d_{i,t+1}N_{i-1,r+1}Z_{i-1,k+1}\nonumber \\
    &&-{\overline{\Delta}}_{i,j-k-r-t+1}L_{i,t+1}^*Y_{i-1,r+1}Z_{i-1,k+1},\\
    &&\quad 0 \!\leq\! j \!\leq\! \overline{b}_i=2p_{i-1}+{\overline{\overline{n}}_q+q_{i-1}+\hat{q}_{i-1}+n_q},\nonumber\\
W_{i,j+1}\!\!&=&\!\!\!\sum\limits_{r=0}^{j}\sum\limits_{k=0}^{j-r}{\overline{\overline{\Delta}}_{i,j-k-r+1}Y_{i-1,k+1}Y_{i-1,r+1}},\label{tri14}\\
     && 0\! \leq j \leq {\overline{\overline{b}}_i}=2\hat{q}_{i-1}+n_q+\max\{n_q+\overline{n}_q,\overline{\overline{n}}_q\}+2p_{i-1},\nonumber
\end{eqnarray}
\qquad where
\begin{footnotesize}
\begin{eqnarray}\label{tri15}
\overline{\Delta}_{i,j+1}\!\!\!\!&=&\!\!\!\!{\sum\limits_{r=0}^{j}\sum\limits_{k=0}^{j-r}Y_{i-1,j-k-r+1}Y_{i-1,k+1}\overline{\overline{N}}_{i-1,r+1}s^j},\quad
0 \leq j \leq \overline{\delta}_q={2p_{i-1}+\overline{\overline{n}}_q}\\
\overline{\overline{\Delta}}_{i,j+1}\!\!&=&\!\!\sum\limits_{t=0}^{j}\sum\limits_{r=0}^{j-t}\sum\limits_{k=0}^{j-r-t}(\widehat{n}_{i,j-k-r-t+1}Y_{i-1,k+1}Y_{i-1,r+1}+d_{i,j-k-r-t+1}^*N_{i-1,k+1}d_{i,r+1}\nonumber\\
&&\!\!\!\!-d_{i,j-k-r-t+1}^*L_{i,k+1}Y_{i-1,r+1}-L_{i,j-k-r-t+1}^*d_{i,k+1}Y_{i-1,r+1})\overline{\overline{N}}_{i-1,t+1} \label{tri16}\\
&&\!\!\!\!-L_{i,j-k-r+1}^*\varphi_{i,k+1}Y_{i-1,r+1},\quad 0 \leq j \leq \overline{\overline{\delta}}_q=2\hat{q}_{i-1}+n_q+
                    \max\{n_q+\overline{n}_q,\overline{\overline{n}}_q\}\nonumber
\end{eqnarray}
\end{footnotesize}

\leftskip 0.4cm
Then compute
$$Z_{i,j+1},\ 0\!\leq \!j\!\leq \!q_i,\ \ Y_{i,j+1},\ 0\!\leq \!j\leq \!p_i$$
as it is defined in

\leftskip 0cm
\begin{eqnarray}\label{tri17}
Z_{i,j+1}&=&\left[ \begin{array}{ccccc}
   \Theta_{i,j+1}
          \\ \\
 {\sum\limits_{k=0}^{j}\psi_{i,j-k+1}V_{i,k+1}}
    \end{array}\right], \qquad 0\leq j\leq q_i,\ i\geq 2\\
\label{tri18}
Y_{i,j+1}&=&\sum\limits_{k=0}^{j}\psi_{i,j-k+1}W_{i,k+1},\qquad 0\le j\le p_i,\ i\geq 2,
\end{eqnarray}
\noindent where in $(\ref{tri17})$ and $(\ref{tri18})$ is:
\begin{eqnarray}
\Theta_{i,j+1}\!\!\!\!\!&=&\!\!\!\!\!\sum\limits_{r=0}^{j}\sum\limits_{k=0}^{j-r}
          Z_{i-1,j-k-r+1}\overline{\overline{N}}_{i-1,k+1}W_{i,r+1} \nonumber \\
          && \ -d_{i,j-k-r+1}\overline{\overline{N}}_{i-1,k+1}V_{i,r+1} \!-\!\varphi_{i,j-k+1}V_{i,k+1},  \label{tri19}\\
          && 0\!\le \!j\le \! q_i, \nonumber \\
{}\nonumber \\
\varphi_{i,j+1}\!\!\!\!&=&\!\!\!\!\!\!\!\sum\limits_{t=0}^j\sum\limits_{r=0}^{j-t}\sum\limits_{k=0}^{j-t-r}\!
           \left( Y_{i-1,j-k-r-t+1}I_{i-1,k+1}-Z_{i-1,j-k-r-t+1}\widehat{A}_{i-1,k+1}\right)\nonumber \\
           && \qquad \times {\overline{N}_{i-1,r+1}L_{i,k+1}},\label{tri20}\\
&&0\le j \le \hat{q}_{i-1}+\overline{n}_q+n_q,\nonumber \\
{}\nonumber  \\
\psi_{i,j+1}\!\!\!\!&=&\!\!\!\!\!\sum\limits_{k=0}^{j}Y_{i-1,j-k+1}\overline{\overline{N}}_{i-1,k+1},\quad 0\le j \le {p_{i-1}+
                     \overline{\overline{n}}_q}\label{tri21}
\end{eqnarray}
\noindent and
\begin{eqnarray}\label{tri22}
q_i\!\!\!\!&=&\!\!\!\!\hat{q}_{i-1}+q+\max\{\overline{n}_q+n_q,\overline{\overline{n}}_q\}+\max\{\overline{b}_i,\overline{\overline{b}}_i\},\quad  q_1=q+m_q\\
\label{tri23}
p_i\!\!\!\!&=&\!\!\!\!p_{i-1}+\overline{\overline{n}}_q+\overline{\overline{b}}_i, \quad  p_1=2q+m_q.
\end{eqnarray}

\noindent {\it Step 2.4.\/} Compute
\begin{equation}\label{tri24}
X_{i}(s) =   \frac{\sum\limits_{j=0}^{q_i}Z_{i,j+1}s^j}
                  {\sum\limits_{j=0}^{p_i}Y_{i,j+1}s^j},
\end{equation}
where $q_i$ and $p_i$ are defined in $(\ref{tri22})$ and $(\ref{tri23})$, respectively.

\smallskip
\noindent {\it Step 3.\/} The stopping criterion is $i=n$. In this case the result is the weighted Moore-Penrose inverse
$A(s)_{M,N}^\dagger =X_{n}(s)$.
\end{alg}
\end{thm}

\begin{demo}
If $a_1(s)=0$, in view of the second case in $(\ref{dva3})$ we have
\begin{eqnarray*}
  X_{1}(s)=a_1(s)^\dagger=\sum\limits_{j=0}^{q} a_{1,j+1}^*, \qquad Y_{1}(s)=1.
\end{eqnarray*}
If $a_1(s)\neq 0$, in accordance with the first case in $(\ref{dva3})$ we have

\begin{eqnarray*}
  X_1(s)\!\!\! &=&\!\!\!\! \left(a_1^*(s)M(s)a_1(s)\right)^{-1}a_1^*(s)M(s) \\
   \!\!&=&\!\!\!\! \left( \sum\limits_{j=0}^q a_{1,j+1}^* s^j \sum\limits_{j=0}^{m_q} M_{j+1} s^j \sum\limits_{j=0}^q a_{1,j+1} s^j \right) ^{-1}
                          \sum\limits_{j=0}^q a_{1,j+1}^*s^j \sum\limits_{j=0}^{m_q} M_{j+1} s^j\\
\!\!&=&\!\!\!\! \left(\sum\limits_{j=0}^{2q+m_q} \sum\limits_{r=0}^j \sum\limits_{k=0}^{j-k} a_{1,j-k-r+1}^*M_{k+1}a_{1,r+1} s^j\right) ^{-1}
  \sum\limits_{j=0}^{q+m_q} \sum\limits_{k=0}^{j} a_{1,j-k+1}^*M_{k+1}s^j\\
\!\!&=&\!\!\!\frac{\sum\limits_{j=0}^{q+m_q} \sum\limits_{k=0}^{j} a_{1,j-k+1}^*M_{k+1}s^j}
        {\sum\limits_{j=0}^{2q+m_q}\sum\limits_{r=0}^{j} \sum\limits_{k=0}^{j-r} a_{1,j-k-r+1}^* M_{k+1}a_{1,r+1}s^j}.
\end{eqnarray*}

Therefore, $X_{1}(s)$ is the partial
case $i=1$ of $(\ref{tri24})$, where the matrices $Z_{1,j+1}$ and $Y_{1,j+1}$
are defined in $(\ref{tri6})$ and $(\ref{tri7})$, respectively.

\smallskip
For each $i=2,\ldots ,n$ it is reasonable to calculate matrices
$X_{i}(s)$ in the form (\ref{tri24}),
for appropriate matrices $Z_{i,j+1}$, $Y_{i,j+1}$ and appropriate upper bounds $q_i$ and $p_i$.

\smallskip
Direct calculation in $(\ref{dva5})$, i.e. {\it Step 2.1\/} of {\it Algorithm 2.1\/} yields the following
\begin{eqnarray*}
d_i(s)&=& X_{i-1}(s) a_i(s)=
\frac{\sum\limits_{j=0}^{q_{i-1}}Z_{i-1,j+1}s^j}
      {\sum\limits_{j=0}^{p_{i-1}}Y_{i-1,j+1}s^j} \cdot
\sum\limits_{k=0}^{q}a_{i,k+1}s^k\\
&=&\frac {\sum\limits_{j=0}^{q_{i-1}+q}
         (\sum\limits_{k=0}^j Z_{i-1,j-k+1}a_{i,k+1})s^j}
      {\sum\limits_{j=0}^{p_{i-1}}Y_{i-1,j+1}s^j}.
\end{eqnarray*}
Then $d_i(s)$ can be represented in the form
\begin{eqnarray}\label{tri25}
d_i(s)=\frac{\sum\limits_{j=0}^{q_{i-1}+q}d_{i,j+1}s^j}
       {\sum\limits_{j=0}^{p_{i-1}}Y_{i-1,j+1}s^j},
\end{eqnarray}
where the matrices $d_{i,j+1}$ are defined by (\ref{tri8}).

\smallskip
Consider $(\ref{dva6})$, i.e. {\it Step 2.2\/} of {\it Algorithm 2.1\/}. Since the
first $i-1$ columns of $A(s)$ can be represented in the polynomial form
$$\widehat{A}_{i-1}(s)=\sum\limits _{j=0}^q \widehat{A}_{i-1,j+1} s^j$$
for appropriate $m\times (i-1)$ constant matrices $\widehat{A}_{i-1,j+1}(s)$, in view of (\ref{tri3}) and (\ref{tri25}) we obtain
\begin{eqnarray*}
c_i(s)\!\!\!\! &=&\!\!\!\! a_i(s)\!-\!\widehat{A}_{i-1}(s)d_i(s)\\
\\
\!\!\!\!&=&\!\!\!\! \sum\limits_{j=0}^{q} a_{i,j+1}s^j- \sum\limits_{j=0}^{q} \widehat{A}_{i-1,j+1}s^j \cdot
\frac{\sum\limits_{j=0}^{q_{i-1}+q}d_{i,j+1}s^j}
     {\sum\limits_{j=0}^{p_{i-1}}Y_{i-1,j+1}s^j}\\
     \\
\!\!\!\!&=&\!\!\!\! \frac{\sum\limits_{j=0}^{q+p_{i-1}}
 \left( \sum\limits_{k=0}^j a_{i,j-k+1}Y_{i-1,k+1}\right) s^j-
\sum\limits_{j=0}^{2q+q_{i-1}}
   (\sum\limits_{k=0}^j \widehat{A}_{i-1,j-k+1}d_{i,k+1})s^j}
   {\sum\limits_{j=0}^{p_{i-1}}Y_{i-1,j+1}s^j}
\end{eqnarray*}

Finding a maximum between the upper bounds $q+p_{i-1}$ and $2q+q_{i-1}$ in the last identity, we have
\begin{eqnarray*}
 c_i(s) \!\!\!\!&=&\!\!\!\! \frac{\sum\limits_{j=0}^{\hat{q}_{i-1}+q}\sum\limits_{k=0}^j
(a_{i,j-k+1}Y_{i-1,k+1}-\widehat{A}_{i-1,j-k+1}d_{i,k+1})s^j}
{\sum\limits_{j=0}^{p_{i-1}}Y_{i-1,j+1}s^j}.
\phantom{ \sum\limits_{j=0}^{q_{i-1}+q}d_{i,j+1}s^j},
\end{eqnarray*}
where $\hat{q}_{i-1}$ is defined in $(\ref{tri10})$ and shorter polynomial matrix is filled by appropriate zero matrices.

Therefore, $c_i(s)$ can be represented in the form
\begin{eqnarray*}
c_i(s)=\frac{\sum\limits_{j=0}^{\hat{q}_{i-1}+q}c_{i,j+1}s^j}
       {\sum\limits_{j=0}^{p_{i-1}}Y_{i-1,j+1}s^j},
\end{eqnarray*}
where $c_{i,j+1}$ are matrices of the form $(\ref{tri9})$, for each
$0\!\leq \!j \!\leq \!\hat{q}_{i-1}+q.$

\smallskip
Observe now {\it Step 2.3.\/} of {\it Algorithm 2.1\/}, i.e $(\ref{dva6})$.

\smallskip
If $c_{i,j+1}\neq 0$ for some $j$, then $c_i(s)\ne 0$ and $b_i^*(s)$ is equal to

\begin{eqnarray*}
& b_i^*(s)& =\left( c_i^*(s) M(s) c_i(s)\right) ^{-1} c_i^*(s) M(s)\\
&=&\!\!\!\!\!\!\! \left[\frac{\sum\limits_{j=0}^{\hat{q}_{i-1}+q}c_{i,j+1}^*s^j}
       {\sum\limits_{j=0}^{p_{i-1}}Y_{i-1,j+1}s^j}\sum\limits_{j=0}^{m_q} M_{j+1}s^j
       \frac{\sum\limits_{j=0}^{\hat{q}_{i-1}+q}c_{i,j+1}s^j}
       {\sum\limits_{j=0}^{p_{i-1}}Y_{i-1,j+1}s^j}\right]^{-1}
       \frac{\sum\limits_{j=0}^{\hat{q}_{i-1}+q}c_{i,j+1}^*s^j}
            {\sum\limits_{j=0}^{p_{i-1}}Y_{i-1,j+1}s^j}\sum\limits_{j=0}^{m_q} M_{j+1}s^j\\
       \!\!\!\!&=&\!\!\!\!\frac{\sum\limits_{j=0}^{p_{i-1}}Y_{i-1,j+1}s^j.\sum\limits_{j=0}^{\hat{q}_{i-1}+q+m_q}\sum\limits_{k=0}^{j}c_{i,j-k+1}^*M_{k+1}s^j}
                {\sum\limits_{j=0}^{2\hat{q}_{i-1}+2q+m_q}\sum\limits_{r=0}^{j}\sum\limits_{k=0}^{j-r}c_{i,j-k-r+1}^*M_{k+1}c_{i,r+1}s^j}\\
       \!\!\!\!&=&\!\!\!\!\frac{\sum\limits_{j=0}^{\hat{q}_{i-1}+q+p_{i-1}+m_q}\sum\limits_{r=0}^{j}\sum\limits_{k=0}^{j-r}Y_{i-1,j-k-r+1}c_{i,k+1}^*M_{r+1}s^j}
               {\sum\limits_{j=0}^{2\hat{q}_{i-1}+2q+m_q}\sum\limits_{r=0}^{j}\sum\limits_{k=0}^{j-r}c_{i,j-k-r+1}^*M_{k+1}c_{i,r+1}s^j}
       \!\!=\!\!\frac{\sum\limits_{j=0}^{\overline{b}_i}V_{i,j+1}s^j}
                {\sum\limits_{j=0}^{\overline{\overline{b}}_i}W_{i,j+1}s^j}
\end{eqnarray*}
where $V_{i,j+1}$ and $W_{i,j+1}$ satisfy $(\ref{tri11})$ and $(\ref{tri12})$, respectively.

\smallskip
If $c_{i,j+1}=0$ for all $j$, then $c_i(s)= 0$ and $b_i^*(s)$ is defined in the second case of $(\ref{dva7})$ and in $(\ref{dva8})$.
In order to compute $\delta _i^{-1}(s)$, we firstly generate the following intermediate value, which will be used later:
\begin{scriptsize}
\begin{eqnarray*}
&\sigma_i(s)&\!\!\!\!= \left( I-X_{i-1}(s) \widehat{A}_{i-1}(s)\right) N_{i-1}^{-1}(s) l_i(s)\\
&=&\!\!\!\!\left( I-\frac{\sum\limits_{j=0}^{q_{i-1}}Z_{i-1,j+1}s^j} {\sum\limits_{j=0}^{p_{i-1}}Y_{i-1,j+1}s^j}
    \sum\limits_{j=0}^q \widehat{A}_{i-1,j+1}s^j\right)
                \frac{\sum\limits_{j=0}^{\overline{n}_q} \overline{N}_{i-1,j+1}s^j}{\sum\limits_{j=0}^{\overline{\overline{n}}_q} \overline{\overline{N}}_{i-1,j+1}s^j}
    {\sum\limits_{j=0}^{n_q}L_{i,j+1}s^j}\\
    &=&\!\!\!\!\frac{\sum\limits_{j=0}^{\hat{q}_{i-1}}\sum\limits_{k=0}^jY_{i-1,j-k+1}I_{i-1,k+1}-Z_{i-1,j-k+1}\widehat{A}_{i-1,k+1}s^j}{\sum\limits_{j=0}^{p_{i-1}}Y_{i-1,j+1}s^j}
.\frac{\sum\limits_{j=0}^{\overline{n}_q+n_q}\sum\limits_{k=0}^{j}  \overline{N}_{i-1,j-k+1}L_{i,k+1}s^j}{\sum\limits_{j=0}^{\overline{\overline{n}}_q} \overline{\overline{N}}_{i-1,j+1}s^j}\\
&=&\!\!\!\!\!\!\!\!
      \frac{\sum\limits_{j=0}^{\hat{q}_{i-1}+\overline{n}_q+n_q}{\sum\limits_{t=0}^j\sum\limits_{r=0}^{j-t}\sum\limits_{k=0}^{j-t-r}(Y_{i-1,j-k-r-t+1}I_{i-1,k+1}
                  -Z_{i-1,j-k-r-t+1}\widehat{A}_{i-1,k+1}){\overline{N}_{i-1,r+1}L_{i,k+1}}}}
          {\sum\limits_{j=0}^{p_{i-1}+\overline{\overline{n}}_q}\sum\limits_{k=0}^{j}Y_{i-1,j-k+1}\overline{\overline{N}}_{i-1,k+1}s^j}\\
&=&\!\!\!\!\frac{\sum\limits_{j=0}^{\hat{q}_{i-1}+\overline{n}_q+n_q}\varphi_{i,j+1}s^j}
        {\sum\limits_{j=0}^{p_{i-1}+\overline{\overline{n}}_q}\psi_{i,j+1}s^j}.
\end{eqnarray*}
\end{scriptsize}
\noindent In the last identity $\varphi_{i,j+1}$ and $\psi_{i,j+1}$ and $\delta_i$ are defined by $(\ref{tri20})$ and $(\ref{tri21})$.

\smallskip
Now, $\delta _i(s)$ is equal to
\begin{scriptsize}
\begin{eqnarray*}
&&\delta_i(s)= n_{ii}(s)+d^*_i(s) N_{i-1}(s) d_i(s)-(d^*_i(s)l_i(s)+l_i^*(s)d_i(s))-l_i^*(s)\sigma_i(s)\\
&&=\sum\limits_{j=0}^{n_q}\widehat{n}_{i,j+1}s^j+\frac{\sum\limits_{j=0}^{q_{i-1}+q}d_{i,j+1}^*s^j}{\sum\limits_{j=0}^{p_{i-1}}Y_{i-1,j+1}}
\sum\limits_{j=0}^{n_q}N_{i-1,j+1}s^j \frac{\sum\limits_{j=0}^{q_{i-1}+q}d_{i,j+1}s^j}{\sum\limits_{j=0}^{p_{i-1}}Y_{i-1,j+1}}\\
&&-\frac{\sum\limits_{j=0}^{q_{i-1}+q}d_{i,j+1}^*s^j}{\sum\limits_{j=0}^{p_{i-1}}Y_{i-1,j+1}} \sum\limits_{j=0}^{n_q}L_{i,j+1}s^j
-\sum\limits_{j=0}^{n_q}L_{i,j+1}^*s^j\frac{\sum\limits_{j=0}^{q_{i-1}+q}d_{i,j+1}s^j}{\sum\limits_{j=0}^{p_{i-1}}Y_{i-1,j+1}}
-\sum\limits_{j=0}^{n_q}L_{i,j+1}^*s^j \frac{\sum\limits_{j=0}^{\hat{q}_{i-1}+\overline{n}_q+n_q}\varphi_{i,j+1}s^j}
        {\sum\limits_{j=0}^{p_{i-1}+\overline{\overline{n}}_q}\psi_{i,j+1}s^j}\\
&&=\frac{\sum\limits_{j=0}^{2\hat{q}_{i-1}+n_q}\sum\limits_{r=0}^{j}\sum\limits_{k=0}^{j-r}\widehat{n}_{i,j-k-r+1}Y_{i-1,k+1}Y_{i-1,r+1}+d_{i,j-k-r+1}^*N_{i-1,k+1}d_{i,r+1}s^j}
        {\sum\limits_{j=0}^{2p_{i-1}}\sum\limits_{k=0}^{j}Y_{i-1,j-k+1}Y_{i-1,k+1}s^j}\\
&&-\frac{\sum\limits_{j=0}^{q_{i-1}+q+p_{i-1}+n_q}\sum\limits_{r=0}^{j}\sum\limits_{k=0}^{j-r}d_{i,j-k-r+1}^*L_{i,k+1}Y_{i-1,r+1}+L_{i,j-k-r+1}^*d_{i,k+1}Y_{i-1,r+1}s^j}
        {\sum\limits_{j=0}^{2p_{i-1}}\sum\limits_{k=0}^{j}Y_{i-1,j-k+1}Y_{i-1,k+1}s^j}\\
&&-\frac{\sum\limits_{j=0}^{\hat{q}_{i-1}+2n_q+\overline{n}_q}\sum\limits_{k=0}^{j}L_{i,j-k+1}^*\varphi_{i,k+1}s^j}
        {\sum\limits_{j=0}^{p_{i-1}+\overline{\overline{n}}_q}\psi_{i,j+1}s^j}
\\
&&=\frac{\sum\limits_{j=0}^{2\hat{q}_{i-1}+n_q+\overline{\overline{n}}_q}\sum\limits_{t=0}^{j}\sum\limits_{r=0}^{j-t}\sum\limits_{k=0}^{j-r-t}(\widehat{n}_{i,j-k-r-t+1}Y_{i-1,k+1}Y_{i-1,r+1}+d_{i,j-k-r-t+1}^*N_{i-1,k+1}d_{i,r+1})\overline{\overline{N}}_{i-1,t+1}s^j}
{\sum\limits_{j=0}^{2p_{i-1}+\overline{\overline{n}}_q}\sum\limits_{r=0}^{j}\sum\limits_{k=0}^{j-r}Y_{i-1,j-k-r+1}Y_{i-1,k+1}\overline{\overline{N}}_{i-1,r+1}s^j}\\
&&-\frac{\sum\limits_{j=0}^{2\hat{q}_{i-1}+n_q+\overline{\overline{n}}_q}\sum\limits_{t=0}^{j}\sum\limits_{r=0}^{j-t}\sum\limits_{k=0}^{j-r-t}(d_{i,j-k-r-t+1}^*L_{i,k+1}Y_{i-1,r+1}+L_{i,j-k-r-t+1}^*d_{i,k+1}Y_{i-1,r+1})\overline{\overline{N}}_{i-1,t+1}s^j}
{\sum\limits_{j=0}^{2p_{i-1}+\overline{\overline{n}}_q}\sum\limits_{r=0}^{j}\sum\limits_{k=0}^{j-r}Y_{i-1,j-k-r+1}Y_{i-1,k+1}\overline{\overline{N}}_{i-1,r+1}s^j}\\
&&-\frac{\sum\limits_{j=0}^{\hat{q}_{i-1}+2n_q+p_{i-1}+\overline{n}_q}\sum\limits_{r=0}^{j}\sum\limits_{k=0}^{j-r}L_{i,j-k-r+1}^*\varphi_{i,k+1}Y_{i-1,r+1}s^j}
{\sum\limits_{j=0}^{2p_{i-1}+\overline{\overline{n}}_q}\sum\limits_{r=0}^{j}\sum\limits_{k=0}^{j-r}Y_{i-1,j-k-r+1}Y_{i-1,k+1}\overline{\overline{N}}_{i-1,r+1}s^j}
\\
&&=\frac{\sum\limits_{j=0}^{\overline{\overline{\delta}}_q}\sum\limits_{t=0}^{j}\sum\limits_{r=0}^{j-t}\sum\limits_{k=0}^{j-r-t}(\widehat{n}_{i,j-k-r-t+1}Y_{i-1,k+1}Y_{i-1,r+1}+d_{i,j-k-r-t+1}^*N_{i-1,k+1}d_{i,r+1})\overline{\overline{N}}_{i-1,t+1}s^j}
{\sum\limits_{j=0}^{2p_{i-1}+\overline{\overline{n}}_q}\sum\limits_{r=0}^{j}\sum\limits_{k=0}^{j-r}Y_{i-1,j-k-r+1}Y_{i-1,k+1}\overline{\overline{N}}_{i-1,r+1}s^j}\\
&&-\frac{\sum\limits_{j=0}^{\overline{\overline{\delta}}_q}\sum\limits_{t=0}^{j}\sum\limits_{r=0}^{j-t}\sum\limits_{k=0}^{j-r-t}(d_{i,j-k-r-t+1}^*L_{i,k+1}Y_{i-1,r+1}+L_{i,j-k-r-t+1}^*d_{i,k+1}Y_{i-1,r+1})\overline{\overline{N}}_{i-1,t+1}s^j}
{\sum\limits_{j=0}^{2p_{i-1}+\overline{\overline{n}}_q}\sum\limits_{r=0}^{j}\sum\limits_{k=0}^{j-r}Y_{i-1,j-k-r+1}Y_{i-1,k+1}\overline{\overline{N}}_{i-1,r+1}s^j}\\
&&-\frac{\sum\limits_{j=0}^{\overline{\overline{\delta}}_q}\sum\limits_{r=0}^{j}\sum\limits_{k=0}^{j-r}L_{i,j-k-r+1}^*\varphi_{i,k+1}Y_{i-1,r+1}s^j}
{\sum\limits_{j=0}^{2p_{i-1}+\overline{\overline{n}}_q}\sum\limits_{r=0}^{j}\sum\limits_{k=0}^{j-r}Y_{i-1,j-k-r+1}Y_{i-1,k+1}\overline{\overline{N}}_{i-1,r+1}s^j}
=\frac{\sum\limits_{j=0}^{\overline{\overline{\delta}}_q}{\overline{\overline{\Delta}}_{i,j+1}}s^j}
{\sum\limits_{j=0}^{\overline{\delta}_q}\overline{\Delta}_{i,j+1}s^j}.
\end{eqnarray*}
\end{scriptsize}
Therefore
$$\delta_i(s)^{-1}=\frac{\sum\limits_{j=0}^{\overline{\delta}_q}\overline{\Delta}_{i,j+1}s^j}
        {\sum\limits_{j=0}^{\overline{\overline{\delta}}_q}{\overline{\overline{\Delta}}_{i,j+1}}s^j},$$
where $\overline{\Delta}_{i,j+1}$ and $\overline{\overline{\Delta}}_{i,j+1}$ are defined in $(\ref{tri15})$ and $(\ref{tri16})$, respectively.

\smallskip
Now, in accordance with the second case of $(\ref{dva7})$, $b_i(s)$ is equal to

\begin{footnotesize}
\begin{eqnarray*}
&& b_i^*(s)=\delta_i^{-1}(s)\left( d_i^*(s) N_{i-1}(s)-l_i^*(s)\right) X_{i-1}(s)\\
&&=\frac{\sum\limits_{j=0}^{\overline{\delta}_q}\overline{\Delta}_{i,j+1}s^j}
        {\sum\limits_{j=0}^{\overline{\overline{\delta}}_q}{\overline{\overline{\Delta}}_{i,j+1}}s^j}
        \left(\frac{\sum\limits_{j=0}^{q_{i-1}+q}d_{i,j+1}^*s^j}
       {\sum\limits_{j=0}^{p_{i-1}}Y_{i-1,j+1}s^j}{\sum\limits_{j=0}^{n_q}N_{i-1,j+1}s^j}-{\sum\limits_{j=0}^{n_q} L_{i,j+1}^*s^j}\right)
      \frac{\sum\limits_{j=0}^{q_{i-1}}Z_{i-1,j+1}s^j} {\sum\limits_{j=0}^{p_{i-1}}Y_{i-1,j+1}s^j}\\
&&=\frac{\sum\limits_{j=0}^{\overline{\delta}_q}\overline{\Delta}_{i,j+1}s^j
        .\sum\limits_{j=0}^{\hat{q}_{i-1}+n_q}\left(\sum\limits_{k=0}^{j}d_{i,j-k+1}^*N_{i-1,k+1}-L_{i,j-k+1}^*Y_{i-1,k+1}\right)s^j
        .\sum\limits_{j=0}^{q_{i-1}}Z_{i-1,j+1}s^j}
{\sum\limits_{j=0}^{\overline{\overline{\delta}}_q+2p_{i-1}}
    \left(\sum\limits_{r=0}^{j}\sum\limits_{k=0}^{j-r}{\overline{\overline{\Delta}}_{i,j-k-r+1}Y_{i-1,k+1}Y_{i-1,r+1}}\right)s^j}\\
&&=\frac{\sum\limits_{j=0}^{\overline{\delta}_q+q_{i-1}+\hat{q}_{i-1}+n_q}
    \sum\limits_{t=0}^{j}\sum\limits_{r=0}^{j-t}\sum\limits_{k=0}^{j-t-r}{\overline{\Delta}}_{i,j-k-r-t+1}d_{i,k+1}^*N_{i-1,r+1}Z_{i-1,t+1}s^j}
   {\sum\limits_{j=0}^{\overline{\overline{\delta}}_q+2p_{i-1}}
    \left(\sum\limits_{r=0}^{j}\sum\limits_{k=0}^{j-r}{\overline{\overline{\Delta}}_{i,j-k-r+1}Y_{i-1,k+1}Y_{i-1,r+1}}\right)s^j}\\
&\qquad &\quad -\frac{\sum\limits_{j=0}^{\overline{\delta}_q+q_{i-1}+\hat{q}_{i-1}+n_q}
    \sum\limits_{t=0}^{j}\sum\limits_{r=0}^{j-t}\sum\limits_{k=0}^{j-t-r}{\overline{\Delta}}_{i,j-k-r-t+1}L_{i,k+1}^*Y_{i-1,r+1}Z_{i-1,t+1}s^j}
   {\sum\limits_{j=0}^{\overline{\overline{\delta}}_q+2p_{i-1}}
    \left(\sum\limits_{r=0}^{j}\sum\limits_{k=0}^{j-r}{\overline{\overline{\Delta}}_{i,j-k-r+1}Y_{i-1,k+1}Y_{i-1,r+1}}\right)s^j}\\
&&=\frac{\sum\limits_{j=0}^{\overline{b}_i}V_{i,j+1}s^j}
                {\sum\limits_{j=0}^{\overline{\overline{b}}_i}W_{i,j+1}s^j}.
\end{eqnarray*}
\end{footnotesize}
\noindent It is not difficult to verify that in the last expression $V_{i,j+1}$ and $W_{i,j+1}$ satisfy $(\ref{tri13})$ and $(\ref{tri14})$, respectively.

\smallskip
Finally, using $(\ref{dva4})$ of {\it Algorithm 2.1\/}, we obtain

\begin{small}
\begin{eqnarray*}
X_{i}(s)\!\!\!\!\!&=&\!\!\!\!\!\left[ \begin{array}{ccc}
      X_{i-1}(s)-\left( d_i(s)+\sigma_i(s)\right) b_i^*(s)
      \\
       b_i^*(s)
    \end{array}\right]\\
\!\!\!\!\!&=&\!\!\!\!\!\left[ \begin{array}{cccc}
   \frac{\sum\limits_{j=0}^{q_{i-1}}Z_{i-1,j+1}s^j}
        {\sum\limits_{j=0}^{p_{i-1}}Y_{i-1,j+1}s^j}
        -\left( \frac{\sum\limits_{j=0}^{q_{i-1}+q}d_{i,j+1}s^j}
               {\sum\limits_{j=0}^{p_{i-1}}Y_{i-1,j+1}s^j}
         +\frac{\sum\limits_{j=0}^{\hat{q}_{i-1}+\overline{n}_q+n_q}\varphi_{i,j+1}s^j}
               {\sum\limits_{j=0}^{p_{i-1}+\overline{\overline{n}}_q}\psi_{i,j+1}s^j}\right)
          \frac{\sum\limits_{j=0}^{\overline{b}_i}V_{i,j+1}s^j}
                {\sum\limits_{j=0}^{\overline{\overline{b}}_i}W_{i,j+1}s^j}
 \\
    \frac{\sum\limits_{j=0}^{\overline{b}_i}V_{i,j+1}s^j}
                {\sum\limits_{j=0}^{\overline{\overline{b}}_i}W_{i,j+1}s^j}\\
    \end{array}\right]
  \end{eqnarray*}
\end{small}
\begin{tiny}
 \begin{eqnarray*}
\!\!\!\!\!&&=\left[ \begin{array}{cccc}
   \frac{\sum\limits_{j=0}^{q_{i-1}}Z_{i-1,j+1}s^j}
        {\sum\limits_{j=0}^{p_{i-1}}Y_{i-1,j+1}s^j}
        -\frac{\sum\limits_{j=0}^{q_{i-1}+q+\overline{\overline{n}}_q}\sum\limits_{k=0}^{j}d_{i,j-k+1}\overline{\overline{N}}_{i-1,k+1}s^j
         +{\sum\limits_{j=0}^{\hat{q}_{i-1}+\overline{n}_q+n_q}\varphi_{i,j+1}s^j}}
         {\sum\limits_{j=0}^{p_{i-1}+\overline{\overline{n}}_q}\psi_{i,k+1}s^j}
         .
 \frac{\sum\limits_{j=0}^{\overline{b}_i}V_{i,j+1}s^j}
                {\sum\limits_{j=0}^{\overline{\overline{b}}_i}W_{i,j+1}s^j}
 \\
    \frac{\sum\limits_{j=0}^{\overline{b}_i}V_{i,j+1}s^j}
                {\sum\limits_{j=0}^{\overline{\overline{b}}_i}W_{i,j+1}s^j}\\
    \end{array}\right]
  \end{eqnarray*}
 \begin{eqnarray*}
\!\!\!\!\!&&=\left[ \begin{array}{cccc}
   \frac{\sum\limits_{j=0}^{q_{i-1}}Z_{i-1,j+1}s^j}
        {\sum\limits_{j=0}^{p_{i-1}}Y_{i-1,j+1}s^j}
        -\frac{\left( \sum\limits_{j=0}^{q_{i-1}+q+\overline{\overline{n}}_q}\sum\limits_{k=0}^{j}d_{i,j-k+1}\overline{\overline{N}}_{i-1,k+1}s^j
         +{\sum\limits_{j=0}^{\hat{q}_{i-1}+\overline{n}_q+n_q}\varphi_{i,j+1}s^j}\right) {\sum\limits_{j=0}^{\overline{b}_i}V_{i,j+1}s^j}}
         {\sum\limits_{j=0}^{p_{i-1}+\overline{\overline{n}}_q+\overline{\overline{b}}_i}\sum\limits_{k=0}^{j}\psi_{i,j-k+1}W_{i,k+1}s^j}
 \\
    \frac{\sum\limits_{j=0}^{\overline{b}_i}V_{i,j+1}s^j}
                {\sum\limits_{j=0}^{\overline{\overline{b}}_i}W_{i,j+1}s^j}\\
    \end{array}\right]
\\
\!\!\!\!\!&&=\left[ \begin{array}{cccc}
   \frac{\sum\limits_{j=0}^{\theta_q}\Theta_{i,j+1}s^j}
         {\sum\limits_{j=0}^{p_{i-1}+\overline{\overline{n}}_q+\overline{\overline{b}}_i}\sum\limits_{k=0}^{j}\psi_{i,j-k+1}W_{i,k+1}s^j}
 \\
    \frac{\sum\limits_{j=0}^{\overline{b}_i}V_{i,j+1}s^j}
                {\sum\limits_{j=0}^{\overline{\overline{b}}_i}W_{i,j+1}s^j}\\
    \end{array}\right]=\left[ \begin{array}{cccc}
   \frac{\sum\limits_{j=0}^{\theta_q}\Theta_{i,j+1}s^j}
         {\sum\limits_{j=0}^{p_{i-1}+\overline{\overline{n}}_q+\overline{\overline{b}}_i}\sum\limits_{k=0}^{j}\psi_{i,j-k+1}W_{i,k+1}s^j}
 \\
    \frac{\sum\limits_{j=0}^{\overline{b}_i}V_{i,j+1}s^j}
                {\sum\limits_{j=0}^{\overline{\overline{b}}_i}W_{i,j+1}s^j}.
    \frac{\sum\limits_{j=0}^{p_{i-1}+{\overline{\overline{n}}_q}}\psi_{i,j+1}s^j}
                {\sum\limits_{j=0}^{p_{i-1}+{\overline{\overline{n}}_q}}\psi_{i,j+1}s^j}
    \end{array}\right]\\
\!\!\!\!\!&&=\left[ \begin{array}{cccc}
   \frac{\sum\limits_{j=0}^{\theta_q}\Theta_{i,j+1}s^j}
         {\sum\limits_{j=0}^{p_{i-1}+\overline{\overline{n}}_q+\overline{\overline{b}}_i}\sum\limits_{k=0}^{j}\psi_{i,j-k+1}W_{i,k+1}s^j}
 \\
    \frac{\sum\limits_{j=0}^{p_{i-1}+\overline{\overline{n}}_q+\overline{b}_i}\sum\limits_{k=0}^{j}\psi_{i,j-k+1}V_{i,k+1}s^j}
         {\sum\limits_{j=0}^{p_{i-1}+\overline{\overline{n}}_q+\overline{\overline{b}}_i}\sum\limits_{k=0}^{j}\psi_{i,j-k+1}W_{i,k+1}s^j}
    \end{array}\right]=\frac{\sum\limits_{j=0}^{q_i}Z_{i,j+1}s^j}
          {\sum\limits_{j=0}^{p_i}Y_{i,j+1}s^j},
\end{eqnarray*}
\end{tiny}

\noindent where $\Theta_{i,j+1}$ is defined in $(\ref{tri19})$.

\smallskip
Finally, we obtain the polynomial representations for $Z_{i,j+1}$ and $Y_{i,j+1}$ as in $(\ref{tri17})$-$(\ref{tri23})$

\smallskip
In accordance with Lemma 2.1, the weighted Moore-Penrose inverse for given matrix is $A(s)^{\dagger}_{M,N}=X_{n}(s)$, which completes the proof.
\end{demo}

The next algorithm is a generalization of {\it Algorithm \ref{alg22}} and computes the
inverse matrix $N^{-1}(s)$ in a polynomial form.

\begin{thm}
Let the leading principal submatrix $N_i(s)$ of the positive definite matrix $N(s)$ is partitioned as in $(\ref{dva2})$, and assume that
$n_{ii}(s),l_i(s), N_{i-1}^{-1}(s)$ possesses the polynomial representation $(\ref{tri5})$. Then the following
algorithm computes the inverse matrix $N^{-1}(s)$.

\begin{alg}\label{alginv} Input: positive definite matrix $N(s)$.

\smallskip
\leftskip 0.5cm
{\it Step 1.\/} Initial values:

\leftskip 0cm

\begin{eqnarray}\label{tri26}
\overline{N}_{1,j+1}=1, \quad \overline{\overline{N}}_{1,j+1}=\widehat{n}_{1,j+1}, \quad 0 \leq j \leq \overline{n}_q.
\end{eqnarray}

\leftskip 0.5cm
{\it Step 2.\/} Recursive step: For $2\leq i \leq n$ perform {\it Step 2.1}-{\it Step 2.4}:

\leftskip 0.8cm
\smallskip
{\it Step 2.1.\/} Compute

\leftskip 0cm

\begin{eqnarray}
&&\overline{G}_{i,j+1}=\overline{\overline{N}}_{i-1,j+1}, \quad 0 \leq j \leq \overline{g}_q=\overline{\overline{n}}_q\label{tri27}\\
&&p_{i,j+1}=\sum\limits_{k=0}^{j}\widehat{n}_{i,j-k+1}\overline{\overline{N}}_{i-1,k+1}, \quad 0 \leq j \leq n_q+\overline{n}_q \nonumber\\
&&q_{i,j+1}=\sum\limits_{r=0}^{j}\sum\limits_{k=0}^{j-r}L_{i,j-k-r+1}^*\overline{N}_{i-1,k+1}L_{i,r+1}, \quad 0 \leq j \leq 2n_q+\overline{n}_q \nonumber\\
&&\overline{\overline{G}}_{i,j+1}=p_{i,j+1}-q_{i,j+1},\quad 0 \leq j \leq \overline{\overline{g}}_q=2n_q+\overline{n}_q\label{tri28}
\end{eqnarray}
\leftskip 0cm
\noindent where $p_{i,j+1}$ is padded by zeros from $n_q+\overline{n}_q$ up to upper bound $2n_q+\overline{n}_q$.

\smallskip
\leftskip 0.8cm
{\it Step 2.2.\/} Compute

\leftskip 0cm

\begin{eqnarray}
&&\overline{F}_{i,j+1}=-\sum\limits_{k=0}^j \overline{N}_{i-1,j-k+1}L_{i,k+1}  ,\quad 0 \leq j \leq \overline{f}_q={\overline{n}_q+n_q}\label{tri29}\\
&&\overline{\overline{F}}_{i,j+1}=\overline{\overline{G}}_{i,j+1}, \quad 0 \leq j \leq \overline{\overline{f}}_q=\overline{\overline{g}}_q\label{tri30}
\end{eqnarray}

\leftskip 0.8cm
\item[{\it Step 2.3.\/}] Compute

\leftskip 0cm
\begin{eqnarray}
\overline{E}_{i-1,j+1}&=&\sum\limits_{r=0}^{j}\sum\limits_{k=0}^{j-r}
\overline{N}_{i-1,j-k-r+1}\overline{G}_{i,k+1}\overline{\overline{F}}_{i,r+1}\nonumber\\
&+&\overline{\overline{N}}_{i-1,j-k-r+1}\overline{F}_{i,k+1}\overline{F}_{i-1,r+1}^*,\label{tri31}\\
&&0 \leq j \leq \overline{e}_q=\max(\overline{n}_q+\overline{g}_q+\overline{\overline{f}}_q,\overline{\overline{n}}_q+2\overline{f}_q),\nonumber\\
\overline{\overline{E}}_{i-1,j+1}&=&\overline{\overline{N}}_{i-1,j-k-r+1}\overline{G}_{i,k+1}\overline{\overline{F}}_{i-1,r+1},\label{tri32}\\
&&0 \leq j \leq \overline{\overline{e}}_q=\overline{\overline{n}}_q+\overline{g}_q+\overline{\overline{f}}_q \nonumber.
\end{eqnarray}

\leftskip 0.8cm
\item[{\it Step 2.4.\/}] Generate

\leftskip 0cm

\begin{eqnarray}
&&\overline{N}_{i,j+1}\!\!=\!\!\left[ \!\! \begin{array}{cccc}
{\sum\limits_{r=0}^{j}\sum\limits_{k=0}^{j-r}
{\overline{E}}_{i,j-k-r+1}\overline{\overline{F}}_{i,k+1}\overline{\overline{G}}_{i,r+1}}
&{\sum\limits_{r=0}^{j}\sum\limits_{k=0}^{j-r}
\overline{\overline{E}}_{i,j-k-r+1}{\overline{F}}_{i,k+1}\overline{\overline{G}}_{i,r+1}}\\
{\sum\limits_{r=0}^{j}\sum\limits_{k=0}^{j-r}
\overline{\overline{E}}_{i,j-k-r+1}{\overline{F}}_{i,k+1}^*\overline{\overline{G}}_{i,r+1}}
&{\sum\limits_{r=0}^{j}\sum\limits_{k=0}^{j-r}
\overline{\overline{E}}_{i,j-k-r+1}\overline{\overline{F}}_{i,k+1}{\overline{G}}_{i,r+1}}
\end{array}\!\! \right] \nonumber \\
&& \label{tri33} \\
&& \  0 \leq j \leq \overline{n}_q=\max\{ {\overline{\overline{g}}_q+\overline{\overline{f}}_q+{\overline{e}}_q}
,\ {\overline{\overline{g}}_q+{\overline{f}}_q+\overline{\overline{e}}_q}
,\ {\overline{\overline{g}}_q+{\overline{f}}_q+\overline{\overline{e}}_q}
,\ {{\overline{g}}_q+\overline{\overline{f}}_q+\overline{\overline{e}}_q}\} \nonumber
   \\
   \nonumber \\
&&\overline{\overline{N}}_{i,j+1}=\sum\limits_{r=0}^{j}\sum\limits_{k=0}^{j-r}
\overline{\overline{E}}_{i,j-k-r+1}\overline{\overline{F}}_{i,k+1}\overline{\overline{G}}_{i,r+1}, \nonumber\\
&& \label{tri34} \\
&&\qquad \qquad \qquad 0 \leq j \leq \overline{\overline{n}}_q=\overline{\overline{g}}_q+\overline{\overline{f}}_q+\overline{\overline{e}}_q \nonumber.
\end{eqnarray}

\leftskip 0.5cm
{\it Step 3.\/} Stopping criterion: for $i=n$ the inverse $N^{-1}(s)=N_{n}^{-1}(s)$ is equal to
\begin{eqnarray}\label{tri35}
N^{-1}(s)=\frac{\sum\limits_{j=0}^{\overline{n}_q} \overline{N}_{n,j+1}s^j}
                       {\sum\limits_{j=0}^{\overline{\overline{n}}_q} \overline{\overline{N}}_{n,j+1}s^j}.
\end{eqnarray}

\end{alg}
\end{thm}

\begin{demo}
It is not difficult to verify that $(\ref{tri26})$ follows from
$$N_{1}^{-1}(s)=n_{11}^{-1}(s)=\frac{1}{\sum\limits_{j=0}^{n_q} \widehat{n}_{i,j+1}s^j}.$$

\noindent Also, $(\ref{tri27})$, $(\ref{tri28})$, $(\ref{tri29})$, $(\ref{tri30})$, $(\ref{tri31})$ and $(\ref{tri32})$ follows from
the following.

\noindent Using $(\ref{dva15})$ we have

\begin{footnotesize}
\begin{eqnarray*}
g_{ii}(s)&=&({\sum\limits_{j=0}^{\overline{g}_q} \overline{G}_{i,j+1}s^j})/
                       ({\sum\limits_{j=0}^{\overline{\overline{g}}_q} \overline{\overline{G}}_{i,j+1}s^j})
          =(n_{ii}(s)-l_i^*(s)N_{i-1}^{-1}(s)l_i(s))^{-1}\\
&=&\left(\sum\limits_{j=0}^{n_q} \widehat{n}_{i,j+1}s^j-\sum\limits_{j=0}^{n_q} L_{i,j+1}^*s^j
         \frac{\sum\limits_{j=0}^{\overline{n}_q} \overline{N}_{i-1,j+1}s^j}
              {\sum\limits_{j=0}^{\overline{\overline{n}}_q} \overline{\overline{N}}_{i-1,j+1}s^j}
              \sum\limits_{j=0}^{n_q} L_{i,j+1}s^j\right)^{-1}\\
&=&\frac{\sum\limits_{j=0}^{\overline{\overline{n}}_q} \overline{\overline{N}}_{i-1,j+1}s^j}
{\sum\limits_{j=0}^{n_q+\overline{n}_q}\sum\limits_{k=0}^{j}\widehat{n}_{i,j-k+1}\overline{\overline{N}}_{i-1,k+1}s^j
-\sum\limits_{j=0}^{2n_q+\overline{n}_q}\sum\limits_{r=0}^{j}\sum\limits_{k=0}^{j-r}L_{i,j-k-r+1}^*\overline{N}_{i-1,k+1}L_{i,r+1}s^j}.
\end{eqnarray*}
\end{footnotesize}

An application of $(\ref{dva16})$ gives
\begin{footnotesize}
\begin{eqnarray*}
f_i(s)\!\!\!\!&=&\!\!\!\!({\sum\limits_{j=0}^{\overline{f}_q} \overline{F}_{i,j+1}s^j})/
                       ({\sum\limits_{j=0}^{\overline{\overline{f}}_q} \overline{\overline{F}}_{i,j+1}s^j})
   =-g_{ii}(s)N_{i-1}^{-1}(s)l_i(s)\\
\!\!\!\!&=&\!\!\!\!\frac{-\sum\limits_{j=0}^{\overline{\overline{n}}_q} \overline{G}_{i,j+1}s^j}
                       {\sum\limits_{j=0}^{2n_q+\overline{n}_q} \overline{\overline{G}}_{i,j+1}s^j}
    \frac{\sum\limits_{j=0}^{\overline{n}_q} \overline{N}_{i-1,j+1}s^j}
         {\sum\limits_{j=0}^{\overline{\overline{n}}_q} \overline{\overline{N}}_{i-1,j+1}s^j}
    \sum\limits_{j=0}^{n_q} L_{i,j+1}s^j
=\frac{-\sum\limits_{j=0}^{\overline{n}_q+n_q} \sum\limits_{k=0}^{j}\overline{N}_{i-1,j-k+1}L_{i,k+1}s^j}
         {\sum\limits_{j=0}^{2n_q+\overline{n}_q} \overline{\overline{G}}_{i-1,j+1}s^j}.
\end{eqnarray*}
\end{footnotesize}

In view of $(\ref{dva17})$ one can verify the following:
\begin{footnotesize}
\begin{eqnarray*}
E_{i-1}(s)\!\!\!\!&=&\!\!\!\!({\sum\limits_{j=0}^{\overline{e}_q} \overline{E}_{i,j+1}s^j})/
                       ({\sum\limits_{j=0}^{\overline{\overline{e}}_q} \overline{\overline{E}}_{i,j+1}s^j})
   =N_{i-1}^{-1}(s)+g_{ii}^{-1}(s)f_i(s)f_i^*(s)\\
&\!\!\!\!=&\!\!\!\!\frac{\sum\limits_{j=0}^{\overline{n}_q} \overline{N}_{i-1,j+1}s^j}
         {\sum\limits_{j=0}^{\overline{\overline{n}}_q} \overline{\overline{N}}_{i-1,j+1}s^j}+
         \frac{\sum\limits_{j=0}^{\overline{\overline{g}}_q} \overline{\overline{G}}_{i,j+1}s^j}
         {\sum\limits_{j=0}^{\overline{g}_q} \overline{G}_{i,j+1}s^j}
         \frac{\sum\limits_{j=0}^{\overline{f}_q} \overline{F}_{i,j+1}s^j}
                       {\sum\limits_{j=0}^{\overline{\overline{f}}_q} \overline{\overline{F}}_{i,j+1}s^j}
         \frac{\sum\limits_{j=0}^{\overline{f}_q} \overline{F}_{i,j+1}^*s^j}
                       {\sum\limits_{j=0}^{\overline{\overline{f}}_q} \overline{\overline{F}}_{i,j+1}s^j}\\
&=&\!\!\!\!\frac{\sum\limits_{j=0}^{\max(\overline{n}_q+\overline{g}_q+2\overline{\overline{g}}_q,\overline{\overline{n}}_q+\overline{\overline{g}}_q+2\overline{g}_q)}
\sum\limits_{r=0}^{j}\sum\limits_{k=0}^{j-r}\overline{N}_{i-1,j-k-r+1}\overline{G}_{i,k+1}\overline{\overline{F}}_{i,r+1}}
{\sum\limits_{j=0}^{\overline{\overline{n}}_q+\overline{g}_q+\overline{\overline{f}}_q}\sum\limits_{r=0}^{j}\sum\limits_{k=0}^{j-r}
             \overline{\overline{N}}_{i-1,j-k-r+1}\overline{G}_{i,k+1}\overline{\overline{F}}_{i,r+1}s^j}\\
\qquad &+&\!\!\!\!\frac{\sum\limits_{j=0}^{\max(\overline{n}_q+\overline{g}_q+2\overline{\overline{g}}_q,\overline{\overline{n}}_q+\overline{\overline{g}}_q+2\overline{g}_q)}
\overline{\overline{N}}_{i-1,j-k-r+1}\overline{F}_{i,k+1}^*\overline{F}_{i,r+1}s^j}
        {\sum\limits_{j=0}^{\overline{\overline{n}}_q+\overline{g}_q+\overline{\overline{f}}_q}\sum\limits_{r=0}^{j}\sum\limits_{k=0}^{j-r}
             \overline{\overline{N}}_{i-1,j-k-r+1}\overline{G}_{i,k+1}\overline{\overline{F}}_{i,r+1}s^j}.
\end{eqnarray*}
\end{footnotesize}
Using $(\ref{dva2})$ we finally get the inverse
\begin{footnotesize}
\begin{eqnarray*}
&&N_i^{-1}(s)=\left[
       \begin{array}{ll}
             E_{i-1}(s) & f_i(s) \\
             f_i^*(s) & g_{ii}(s)
       \end{array}
     \right]=\left[
       \begin{array}{ll}
\frac{\sum\limits_{j=0}^{\overline{e}_q} \overline{E}_{i,j+1}s^j}
                       {\sum\limits_{j=0}^{\overline{\overline{e}}_q} \overline{\overline{E}}_{i,j+1}s^j} &
\frac{\sum\limits_{j=0}^{\overline{f}_q} \overline{F}_{i,j+1}s^j}
                       {\sum\limits_{j=0}^{\overline{\overline{f}}_q} \overline{\overline{F}}_{i,j+1}s^j}\\
\frac{\sum\limits_{j=0}^{\overline{f}_q} \overline{F}_{i,j+1}^*s^j}
                       {\sum\limits_{j=0}^{\overline{\overline{f}}_q} \overline{\overline{F}}_{i,j+1}s^j} &
\frac{\sum\limits_{j=0}^{\overline{\overline{n}}_q} \overline{G}_{i,j+1}s^j}
                       {\sum\limits_{j=0}^{2n_q+\overline{n}_q} \overline{\overline{G}}_{i,j+1}s^j}
       \end{array}
     \right]\\
\!\!\!&&=\!\!\!\left[ \begin{array}{cccc}
\frac{\sum\limits_{j=0}^{\overline{\overline{g}}_q+\overline{\overline{f}}_q+\overline{e}_q}\sum\limits_{r=0}^{j}\sum\limits_{k=0}^{j-r}
{\overline{E}}_{i,j-k-r+1}\overline{\overline{F}}_{i,k+1}\overline{\overline{G}}_{i,r+1}}
{\sum\limits_{j=0}^{\overline{\overline{g}}_q+\overline{\overline{f}}_q+\overline{\overline{e}}_q}\sum\limits_{r=0}^{j}\sum\limits_{k=0}^{j-r}
\overline{\overline{E}}_{i,j-k-r+1}\overline{\overline{F}}_{i,k+1}\overline{\overline{G}}_{i,r+1}} &
\frac{\sum\limits_{j=0}^{\overline{\overline{g}}_q+\overline{f}_q+\overline{\overline{e}}_q}\sum\limits_{r=0}^{j}\sum\limits_{k=0}^{j-r}
\overline{\overline{E}}_{i,j-k-r+1}{\overline{F}}_{i,k+1}\overline{\overline{G}}_{i,r+1}}
{\sum\limits_{j=0}^{\overline{\overline{g}}_q+\overline{\overline{f}}_q+\overline{\overline{e}}_q}\sum\limits_{r=0}^{j}\sum\limits_{k=0}^{j-r}
\overline{\overline{E}}_{i,j-k-r+1}\overline{\overline{F}}_{i,k+1}\overline{\overline{G}}_{i,r+1}}\\
\frac{\sum\limits_{j=0}^{\overline{\overline{g}}_q+\overline{f}_q+\overline{\overline{e}}_q}\sum\limits_{r=0}^{j}\sum\limits_{k=0}^{j-r}
\overline{\overline{E}}_{i,j-k-r+1}{\overline{F}}_{i,k+1}^*\overline{\overline{G}}_{i,r+1}}
{\sum\limits_{j=0}^{\overline{\overline{g}}_q+\overline{\overline{f}}_q+\overline{\overline{e}}_q}\sum\limits_{r=0}^{j}\sum\limits_{k=0}^{j-r}
\overline{\overline{E}}_{i,j-k-r+1}\overline{\overline{F}}_{i,k+1}\overline{\overline{G}}_{i,r+1}}&
\frac{\sum\limits_{j=0}^{\overline{g}_q+\overline{\overline{f}}_q+\overline{\overline{e}}_q}\sum\limits_{r=0}^{j}\sum\limits_{k=0}^{j-r}
\overline{\overline{E}}_{i,j-k-r+1}\overline{\overline{F}}_{i,k+1}{\overline{G}}_{i,r+1}}
{\sum\limits_{j=0}^{\overline{\overline{g}}_q+\overline{\overline{f}}_q+\overline{\overline{e}}_q}\sum\limits_{r=0}^{j}\sum\limits_{k=0}^{j-r}
\overline{\overline{E}}_{i,j-k-r+1}\overline{\overline{F}}_{i,k+1}\overline{\overline{G}}_{i,r+1}}
\end{array}\right]
\end{eqnarray*}
\end{footnotesize}
which confirms $(\ref{tri33})$, $(\ref{tri34})$ and $(\ref{tri35})$.
\end{demo}

\section{Examples} 

\begin{exm} 
Find the weighted Moore-Penrose inverse of the rational matrix
\begin{footnotesize}
\begin{verbatim}
X(s)={{s+1,s+2,s},{s,s,s+1},{s+1,s+2,s}}
\end{verbatim}
\end{footnotesize}
\noindent using the following weighting matrices, $M_1(s)$ and $N_1(s)$:
\begin{footnotesize}
\begin{verbatim}
M1(s)={{s+1,s,s+1},{s,s+2,s},{s+1,s,s+3}};
N1(s)={{s+1,s+1,s+1},{s+1,s+2,s},{s+1,s,s+3}};
\end{verbatim}
\end{footnotesize}
The following result is generated applying the function {\it WPartit}, implementing {\it Algorithm 2.1} $($see implementation details$)$:

\smallskip
\begin{footnotesize}
{\tt WPartit[X,M1,N1]}

\smallskip
{\tt WEIGHTED  MOORE-PENROSE  INVERSE=}

$\left(\begin{array}[c]{ccc}
    \frac{2  {s^2}  (2+s)}{12+32  s+33  {s^2}+14  {s^3}}&-\frac{{{(2+s)}^2}}{2+3  s+2  {s^2}}&\frac{s
\big(12+16  s+5  {s^2}\big)}{12+32  s+33  {s^2}+14  {s^3}} \\
    \frac{2+5  s+2  {s^2}}{(6+7  s)  \big(2+3  s+2  {s^2}\big)}&\frac{2  (1+s)}{2+3  s+2  {s^2}}&\frac{4+2
s-2  {s^2}}{12+32  s+33  {s^2}+14  {s^3}} \\
    -\frac{2  {s^2}  (1+s)}{(6+7  s)  \big(2+3  s+2  {s^2}\big)}&\frac{(1+s)  (2+s)}{2+3  s+2
{s^2}}&-\frac{s  (1+s)  (6+5  s)}{(6+7  s)  \big(2+3 s+2 {s^2}\big)}
  \end{array} \right).$
\end{footnotesize}
\end{exm}

\begin{exm} 
In this example we compute the weighted Moore-Penrose inverse of the rational matrix $X(s)$
due to the following weights $M_1(s)$ and $N_1(s)$:
\begin{footnotesize}
\begin{verbatim}
X={{1/s^2,s,(s+1)/s^3},{s,s^2-1,s},{s+1,1/s,s+1}};
M1={{s+1,s,s+1},{s,s+2,s},{s+1,s,s+3}};
N1={{s+1,s+1,s+1},{s+1,s+2,s},{s+1,s,s+3}};
\end{verbatim}

{\tt WPartit[X,M1,N1]}

\smallskip
{\tt WEIGHTED  MOORE-PENROSE  INVERSE=}

\smallskip
$\left(
\begin{array}{rrr}
    -{s^3}&\frac{-1-s+{s^5}+{s^6}}{s \big(-2-s+{s^2}+{s^3}\big)}&\frac{-1-s+{s^2}+{s^3}-{s^5}}{-2-s+{s^2}+{s^3}} \\
    0&\frac{1+s}{-2-s+{s^2}+{s^3}}&\frac{s}{2+s-{s^2}-{s^3}} \\
    {s^3}&-\frac{-1+{s^4}+{s^5}}{-2-s+{s^2}+{s^3}}&\frac{s-{s^3}+{s^5}}{-2-s+{s^2}+{s^3}}
  \end{array} \right).$
\end{footnotesize}
\end{exm}

\begin{exm} 
If the matrices are considered in the polynomial form, then the function {\tt WPartPoly}, implementing {\it Algorithm 3.1}, can be used
to compute the weighted Moore-Penrose inverse of the matrix $X$ $($see implementation details$)$:
\begin{footnotesize}
\begin{verbatim}
X={{1+s,-2+s^4,s},{s,-1+s,s},{s,s,1+s}};  M1=N1={{1+s,s,s},{s,-1+s,s},{s,s,1+s}};
\end{verbatim}
\end{footnotesize}

\begin{footnotesize}
{\tt \noindent WPartPoly[X,M1,N1]}

{\tt WEIGHTED  MOORE-PENROSE  INVERSE=}

$\left(\begin{array}{rrr}
  \frac{1}{1 - s - s^2 + s^5}&\frac{2 + 2\,s + s^2 - s^4 - s^5}{-1 + s + s^2 - s^5}&\frac{s + s^2 - s^5}{1 - s - s^2 + s^5} \\
  \frac{s}{1 - s - s^2 + s^5}&\frac{1 + 2\,s}{-1 + s + s^2 - s^5}&\frac{s}{1 - s - s^2 + s^5}\\
  \frac{s}{-1 + s + s^2 - s^5}&\frac{s\,\left( 3 + s - s^4 \right) }{1 - s - s^2 + s^5}&\frac{-1 + 2\,s + s^2 - s^5}{-1 + s + s^2 - s^5}
\end{array} \right).$
\end{footnotesize}
\end{exm}

\begin{exm}
In this example we generate the Moore-Penrose inverse of the matrix $X(s)$, known as the parameter test matrix of Hessenberg form
$\cite{Zielke}$:

\begin{footnotesize}
\begin{verbatim}
X={{s,1,0,0,0},{s^2,s,1,0,0},{s^3,s^2,s,1,0},{s^4,s^3,s^2,s,1},{s^5,s^4,s^3,s^2,s}}
\end{verbatim}
\end{footnotesize}
\noindent using identity matrices $M1(s)$ and $N1(s)$ of appropriate orders, we get:

\begin{footnotesize}
{\tt \noindent WPartPoly[X,IdentityMatrix[5],IdentityMatrix[5]]}

{\tt WEIGHTED  MOORE-PENROSE  INVERSE=}

$\left(\begin{array}{rrrrr}
          \frac{s}{(1 + s)^2}& 0 &0 &0 &0\\
          \frac{1}{(1 + s)^2} & 0 &0 &0 &0\\
          -s &1 &0 &0 &0\\
          0 &-s &1 &0 &0\\
          0 &0 &-s &\frac{1}{(1 + s)^2}& \frac{s}{(1 + s)^2}
\end{array} \right).$
\end{footnotesize}
\end{exm}

\section{Conclusion}

We extend Wang's partition method from \cite{Wang} to the set of
one-variable rational and polynomial matrices.
In this way, we obtain an algorithm for symbolic computation of the weighted Moore-Penrose inverse
of one-variable rational and polynomial matrices.
The paper is a generalization of the paper \cite{Wang} and a continuation of the paper \cite{Stanimirovic2}.
Several symbolic examples are arranged. In partial case $M=I_m$, $N=I_n$
we obtain the usual Moore-Penrose inverse, and then use test examples from \cite{Zielke}.
Main implementation details are described as the appendix in the next section.

\section{Implementation details}

For the sake of completeness we describe the {\ssr MATHEMATICA} code which
implements Algorithm 2.1. and Algorithm 3.1.

\subsection{Rational matrix case}

Main problem in the implementation of Algorithm 2.1 is the simplification of algebraic expressions included.
This difficulty imposes its implementation in a symbolic computational package.
Moreover, a significant problem in the implementation of {\it Algorithm 2.1\/} is the magnification of
arithmetic operations. This problem increased by multiplicative recomputations.
In view of {\it Step 2\/} in {\it Algorithm 2.1\/}, for each $i\in \{2,\ldots ,n\}$, the Moore-Penrose inverse
$X_{i}(s)$ must be computed $n-i+1$ times.
Moreover, in view of {\it Step 2.1\/} and {\it Step 2.3\/}, the
pseudoinverse $X_{i-1}(s)$ is needful during the computation of the values $d_i(s)$ and $b_i(s)$.
Consequently, {\it Algorithm 2.1\/} requires $3(n-i+1)$ recomputations of the
Moore-Penrose inverse $X_{i}(s)$, for each $i\in \{2,\ldots ,n\}$.
The total number of different values that will
be produced is comparatively small, but these values must be recomputed many times by means of relatively
complicated expressions.
In order to obviate recomputations, we use
possibility of the programming package {\ssr MATHEMATICA\/} to define functions that remember values they have found \cite{Wol,Wol1}.
The pattern for defining a memo function is {\tt f[x\_]:=f[x]=rhs}.

\smallskip
In order to enable simplifications of rational expressions by means of
{\ssr MATHEMATICA\/} function $Simplify$, we restrict our implementation
to the set of rational matrices with real coefficients.

\smallskip
In the beginning we describe two auxiliary procedures.

\smallskip
{\it A.\/} The function $Col[a,j]$ extracts $j$-th column of the matrix $a=A(s)$:
\begin{footnotesize}
\begin{verbatim}
Col[a_List, j_] := Transpose[{Transpose[a][[j]]}]
\end{verbatim}\end{footnotesize}

{\it B.\/} The submatrix $\widehat{A}_j(s)\!=\!\left[a_1(s),\cdots a_j(s)\right]$ which contains first $j\leq n$ columns of the
matrix $A(s)\!=\!\widehat{A}_n(s)\!=\!\left[ a_1(s),\cdots a_n(s)\right]$ is
generated as follows:
\begin{footnotesize}\begin{verbatim}
Adop[a_List,j_]:=Module[{m,n},
  {m,n}=Dimensions[a];
  Return[Transpose[Drop[Transpose[a],-(n-j)]]];]
\end{verbatim}\end{footnotesize}

\smallskip
{\it Step 2\/} of the {\it Algorithm 2.1\/} is implemented in the following
functions which remember before computed values.

\smallskip
Implementation of {\it Step 2.1\/}.

\begin{footnotesize}\begin{verbatim}
DD[a_List,m0_List,n0_List,i_]:=DD[a,m0,n0,i]=
   Module[{s ={}},
    s=Simplify[A[a,m0,n0,i-1].Col[a,i]];
  Return[s]]
\end{verbatim}\end{footnotesize}

Implementation of {\it Step 2.2\/}.
\smallskip
\begin{footnotesize}\begin{verbatim}
CC[a_List,m0_List,n0_List,i_]:=CC[a,m0,n0,i]=
   Module[{s={}},
    s=Col[a,i]-Adop[a,i-1].DD[a,m0,n0,i];
  Return[Simplify[s]]]
\end{verbatim}\end{footnotesize}

Implementation of {\it Step 2.3\/}.

\begin{footnotesize}\begin{verbatim}
B[a_List,m0_List,n0_List,i_]:=B[a,m0,n0,i]=
  Module[{nul,m1,j,k,n1,s={}},
   {m1,n1}=Dimensions[CC[a,m0,n0,i]];
   nul=Table[0,{j,1,m1},{k,1,n1}];
   If[CC[a,m0,n0,i]=!=nul,
     s=Inverse[Transpose[CC[a,m0,n0,i]].m0.CC[a,m0,n0,i]]
      .Transpose[CC[a, m0, n0, i]].m0,
     s=(Delt[a,m0,n0,i])^(-1).(Transpose[DD[a,m0,n0,i]].NK[n0,i][[1]]
                               -Transpose[NK[n0,i][[3]]]).A[a,m0,n0,i-1]];
  Return[Simplify[s]]]
\end{verbatim}\end{footnotesize}

The following function $Delt[a,m0,n0,i]$ computes $\delta_i$ defined in $({\ref{dva8}})$.
\begin{footnotesize}\begin{verbatim}
Delt[a_List,m0_List,n0_List,i_]:=Module[{s},
 s=NK[n0,i][[2]]+Transpose[DD[a,m0,n0,i]].NK[n0,i][[1]].DD[a,m0,n0,i] -
  (Transpose[DD[a,m0,n0,i]].NK[n0,i][[3]]+Transpose[NK[n0,i][[3]]].DD[a,m0,n0,i])
  -Transpose[NK[n0,i][[3]]].(IdentityMatrix[i-1]-A[a,m0,n0,i-1].Adop[a,i-1])
  .Inverse[NK[n0,i][[1]]].NK[n0,i][[3]];
 Return[Simplify[s]]]
\end{verbatim}\end{footnotesize}

In the function $NK[a,i]$ we find the partition $({\ref{dva2}})$ of the leading principal submatrix $N_i(s)$ of the weighted matrix $N(s)$.
\begin{footnotesize}\begin{verbatim}
NK[a_List,i_]:=Module[{lk,NK1,nkk},
  nkk={{a[[i,i]]}};
  If[i==1,Return[{nkk,nkk,nkk}],
     NK1=Transpose[Take[Transpose[Take[a,i-1]],i-1]];
     lk =Transpose[{Most[Last[Take[Transpose[Take[a,i]],i]]]}];
  Return[{NK1,nkk,lk}]]]
\end{verbatim}\end{footnotesize}

Implementation of {\it Step 1\/} and {\it Step 2.4}.

\begin{footnotesize}\begin{verbatim}
A[a_List,m0_List,n0_List,i_]:=A[a,m0,n0,i]=
 Module[{b=a},
 If[i==1,           (* Compute X1(s) *)
   If[Col[a,i]===Col[a,i]*0,
   b=Transpose[a][[1]],                              (* a1(s)=0 *)
   b=Inverse[{Transpose[a][[i]].m0.Col[a,i]}].{Transpose[a][[1]].m0}], (* a1(s)!=0 *)
   (* Compute Xi(s), i>1 *)
 b=A[a,m0,n0,i-1]-(DD[a,m0,n0,i]+(IdentityMatrix[i-1]-A[a,m0,n0,i-1].Adop[a,i-1])
   .Inverse[NK[n0,i][[1]]].NK[n0,i][[3]]).B[a,m0,n0,i];
 b=Append[b,B[a,m0,n0,i][[1]]]];
Return[Simplify[b]]]
\end{verbatim}\end{footnotesize}

The following function starts recursive computations in {\it Step 2}:
\begin{footnotesize}\begin{verbatim}
WPartit[a_List,m0_List,n0_List]:=
  Module[{m,n,i},{m,n}=Dimensions[a];
  Print["WEIGHTED MOORE-PENROSE INVERSE="];
  A[a,m0,n0,n] // MatrixForm]
\end{verbatim}\end{footnotesize}

\subsection{Polynomial matrix case}

We also restrict the implementation to the set of polynomial matrices with real coefficients.
The matrix $A(s)$ defined in $(\ref{tri1})$ can be represented as the list $\{A_1,\ldots ,A_{q+1}\}$.
The $i$-th column $a_i(s)$ of $A(s)$ is the polynomial matrix defined in $(\ref{tri3})$,
and therefore can be represented by the three-dimensional list $\{a_{i,1},\ldots ,a_{i,q+1}\}$, $1\leq i\leq n$.

\begin{scriptsize}
\begin{verbatim}
Col[L_List,j_]:=  (* Compute j-th column from L *)
  Module[{L1=L2={},i},
    For[i=1,i<=Length[L],i++,
        L1=Append[L1,Transpose[L[[i]]]]; AppendTo[L2,Transpose[{L1[[i,j]]}]]];
    Return[L2]];

FrmPoly[M_List]:=   (* Form the polynomial matrix of the form (3.1)  *)
 Module[{L={},i,M1=M,v,s},
   v=Variables[M];
   If[v=!={},
     s=v[[1]];           (* The matrix is not constant *)
     For[i=1, i<=Max[Exponent[M,s]],i++,
        AppendTo[L,Coefficient[M,s^i]];  M1=M1-Coefficient[M,s^i]*s^i];
     M1={M1};
     For[i=1,i<=Length[L],i++, AppendTo[M1,L[[i]]]]];
   If[v=!={},Return[Simplify[M1]],    (* The matrix is not constant *)
             Return[Simplify[{M1}]]]  (* The matrix is constant *)  ];

TakeFPoly[L_List,j_]:=  (* Separate first j columns from L *)
  Block[{L1={},i},
    For[i=1,i<=Length[L],i++,  L1=Append[L1,Take[Transpose[L[[i]]],j]]];
    Return[L1]];

DopZero[L_List,i_]:=    (* Complete the matrix L by zero rows *)
  Module[{L1=L,j,nula},
     nula=L1[[1]]*0;
     For[j=1,j<=i-Length[L],j++,  AppendTo[L1,nula]];
     Return[L1]];

LastZeroP[L_List]:=     (* Drop the last zero rows from L *)
  Module[{L1=L,Us=True,nul,dl},
    If[L1=!={},
       While[Us && L1=!={}, dl=Dimensions[L1][[1]];
         If[L1[[dl]]==L1[[dl]]*0, L1=Drop[L1,-1], Us=False]]];
    Return[L1]];

DDP[L_List,M_List,N_List,i_]:=DDP[L,M,N,i]= (* Compute d_{i,j+1} using (3.8) *)
 Module[{Y={},gr=0,bb,NN,L2={},L1=L,j,nula={}},
   L2=ZZP[L,M,N,i-1];   gr=Length[L]+Length[L2];
   nula={Table[0,{j,1,gr}]};  L1=DopZero[L1,gr];  NN=Col[L1,i]; L2=DopZero[L2,gr];
   For[j=0,j<gr-1,j++,
      If[(j+1)>Length[Y],Y=Join[Y,nula]];
      Y[[j+1]]=Sum[L2[[j-k+1]].NN[[k+1]],{k,0,j}];
   ];
   Y=LastZeroP[Y];Return[Y]];

CCP[L_List,M_List,N_List,i_]:=CCP[L,M,N,i]=  (* Compute c_{i,j+1} using (3.9) *)
 Module[{Y=L4={},gr=0,NN,L1=L,L2=L3={},j,nula={}},
   L2=YYP[L,M,N,i-1];    gr=2Length[L]+Length[L2];  nula=Table[0,{j,1,gr}];
   L1=DopZero[L1,gr]; NN=Col[L1,i];  L2=DopZero[L2,gr]; L4=DDP[L,M,N,i];
   If[L4=={},L4={0}]; L4=DopZero[L4,gr]; L3=TakeFPoly[L,i-1]; L3=DopZero[L3,gr];
   For[j=0,j<gr-1,j++,
     If[(j+1)>Length[Y],Y=Join[Y,nula]];
       If[(Length[L4[[1]]]==0),
         Y[[j+1]]=Sum[NN[[j-k+1]]L2[[k+1]]-(L3[[j-k+1]]L4[[k+1]])[[1]],{k,0,j}],
         Y[[j+1]]=Sum[NN[[j-k+1]]L2[[k+1]]-Transpose[L3[[j-k+1]]].L4[[k+1]],{k,0,j}]
   ]];
   Return[LastZeroP[Y]]];

VVP[L_List,M_List,N_List,i_]:=VVP[L,M,N,i]= (* Compute V_{i,j+1} using (3.11) *)
Module[{M0=M,L1={},L2={},L3={},L4={},L5={},L6={},L7={},
         L8={},Y={},j,k,r,iz,q,mq,q1,gr},
 L2=CCP[L,M,N,i]; L5=DDP[L,M,N,i]; L6=NKP[N,i][[1]];
 L7=NKP[N,i][[3]]; L8=ZZP[L,M,N,i-1];
 If[L2=!={}, L1=YYP[L,M,N,i-1];
   mq=Length[M0]-1; q=Length[L]-1; q1=Length[L1];
   gr=q+2*Length[L1]+mq; L1=DopZero[L1, gr]; L2=DopZero[L2, gr];
   M0=DopZero[M0, gr]; iz = {};
   For[j = 0, j < gr, j++,
    iz=Join[iz,{Sum[Sum[Sum[L1[[j-k-r+1]]Transpose[L2[[k+1]]].
                        M0[[r+1]],{k,0,j-r}],{r,0,j}]][[1]]}]],
 (* Else *)
   iz={};L4=Delt[L,M,N,i][[1]];
   gr=Length[L4]-1+2*Length[XPP[L,M,N,i-1]]+Length[L]-1+Length[N]-1;
   L1=YYP[L,M,N,i-1]; L1 = DopZero[L1, gr]; L4 = DopZero[L4, gr];
   L5 = DopZero[L5, gr]; L6 = DopZero[L6, gr]; L7 = DopZero[L7, gr];
   L8 = DopZero[L8, gr]; Y = {};
   For[j = 0, j < gr, j++,
    If[Length[Dimensions[L5[[1]]]] == 1,
       Y=Join[Y,Sum[L5[[j-k+1]]L6[[k+1]]
             -Transpose[L7[[j-k+1]]]L1[[k+1]], {k,0,j}]],
       Y=Join[Y,Sum[Transpose[L5[[j-k+1]]].L6[[k+1]]
             -Transpose[L7[[j-k+1]]]L1[[k+1]], {k,0,j}]]]];
   Y=DopZero[Y, gr];
   For[j=0,j<gr,j++,
      If[Length[Dimensions[L8[[1]]]] == 1,
      iz = Join[iz,{Sum[Sum[L4[[j-k-r+1,1]]Y[[k+1]].
                   {L8[[r+1]]}, {k,0,j-r}], {r,0,j}]}],
          iz = Join[iz,{Sum[Sum[L4[[j-k-r+1, 1]]Y[[k+1]].
                       L8[[r+1]], {k,0,j-r}], {r,0,j}]}]]]];
   Return[LastZeroP[iz]]];

WWP[L_List,M_List,N_List,i_]:=WWP[L,M,N,i]= (* Compute W_{i,j+1} using (3.12) *)
Module[{Y={},M0=M,gr=0,iz,L0={},L1={},L2={},L3={},L4={},j,nula={},mq},
 L2=CCP[L,M,N,i]; L3=YYP[L,M,N,i-1]; L0=Delt[L,M,N,i][[2]];
 If[L2=!={}, iz={};gr=2*Length[L2]+Length[M0]-1;
  L2=DopZero[L2,gr]; M0=DopZero[M0,gr];
  For[j=0,j<gr,j++,
   iz=Join[iz,Sum[Sum[Transpose[L2[[j-k-r+1]]].M0[[k+1]].L2[[r+1]],{k,0,j-r}],{r,0,j}][[1]]]],
 gr=2*Length[L3]+Length[L0]-1;
 L4=Transpose[{Delt[L,M,N,i][[2]]}];
 iz={};gr=Length[L4]-1+2*Length[L3];
 L3=DopZero[L3,gr]; L4=DopZero[L4,gr];
 For[j=0,j<gr,j++,
   iz=Join[iz,Sum[Sum[L4[[j-k-r+1]]L3[[k+1]]L3[[r+1]],{k,0,j-r}],{r,0,j}]]]];
 Return[LastZeroP[iz]]];

ZZP[L_List,M_List,N_List,i_]:=ZZP[L,M,N,i]=  (* Compute Z_{i,j+1} using (3.17) and (3.6) *)
  Module[{L1=L,L2={},L3={},L4={},L5={},L6={},L7={},L8={},
       M0=M,mq,rez,NN1,q,gr,gr2,iz,iz1,j,k,r},
 If[i==1,  (* Step 1 *)
  mq=Length[M0]-1;q=Length[L1]-1;L2={};
  For[j=1,j<=Length[Col[L1,1]],j++,
      L2=Join[L2,Transpose[Col[L1,1][[j]]]]];
  If[LastZeroP[L2]==={},rez=L2,
    L2=DopZero[L2,mq+q+1]; M0=DopZero[M0,mq+q+1]; iz={};
    For[j=0,j<q+mq+1,j++,
      iz=Join[iz,{Sum[Sum[L2[[j-k+1]].M0[[k+1]],{k,0,j}]]}]];
    rez=LastZeroP[iz];If[rez=={},rez={L2[[1]].M0[[1]]*0}]],
  (*Else *)
    L4=VVP[L,M,N,i]; iz=TET[L,M,N,i]; L7=KSIP[L,M,N,i];
    gr2=Length[iz]+2; iz=DopZero[iz,gr2];
    L4=DopZero[L4,gr2]; L7=DopZero[L7,gr2]; iz1={};
    For[j=0,j<gr2,j++,
       iz1=Join[iz1,{Sum[L7[[j-k+1,1]]L4[[k+1]],{k,0,j}]}]];
    rez={};
    If[LastZeroP[iz]==={},iz=iz1*0];
    For[j=0,j<gr2,j++,
      If[Length[Dimensions[iz[[1]]]] == 1,
        rez=Join[rez,{Join[{iz[[j+1]]},{iz1[[j+1]]}]}],
        If[Dimensions[iz[[1]]][[1]] == 1,
           rez=Join[rez,{Join[iz[[j+1]],{iz1[[j+1]]}]}],
           rez=Join[rez,{Join[iz[[j+1]],{iz1[[j+1]]}]}]
        ]]]];(* EndIF *)
 Return[rez]];

YYP[L_List,M_List,N_List,i_]:=YYP[L,M,N,i]= (* Compute Y_{i,j+1} using (3.18) and (3.7) *)
  Module[{L1=L,L2={},L3={},L4={},L5={},M0=M,iz={},q,mq,j,k,r,gr},
  If[i==1, mq=Length[M0]-1;q=Length[L1]-1;L2={};
    For[j=1,j<=Length[Col[L1,1]],j++,
        L2=Join[L2,Transpose[Col[L1,1][[j]]]]];
    If[LastZeroP[L2]==={},iz=L2,
     L2=DopZero[L2,mq+2*q+1]; M0=DopZero[M0,mq+2*q+1];
     L3=DopZero[Col[L1,1],mq+2*q+1];iz={};
     For[j=0,j<2*q+mq+1,j++,
      iz=Join[iz,Sum[Sum[Sum[L2[[j-k-r+1]].M0[[k+1]].L3[[r+1]],{k,0,j-r}],{r,0,j}]]]];
     iz=LastZeroP[iz]],
  (* Else *)
    L3=KSIP[L,M,N,i];L5=WWP[L,M,N,i];
    gr=Length[L3]+Length[L5]-1;
    L5=DopZero[L5,gr]; L3=DopZero[L3,gr];iz={};
    For[j=0,j<gr,j++,
       iz=Join[iz,Sum[L3[[j-k+1]]L5[[k+1]],{k,0,j}]]];
    iz=LastZeroP[iz]];
  Return[iz]];

NKP[L_List,i_]:=NKP[L,i]=  (* Find the partition (2.2)  *)
 Module[{lk={},NK1={},nkk={},L1={},L2={},L3={},L4={},L5={}},
   For[j=0,j<Length[L],j++,nkk=Join[nkk,{{L[[j+1]][[i,i]]}}]];
   If[i==1,Return[{nkk,nkk,nkk}],
      For[j=0,j<Length[L],j++,L1=Join[L1,{Take[L[[j+1]],i-1]}]];
      For[j=0,j<Length[L],j++,L2=Join[L2,{Transpose[L1[[j+1]]]}]];
      For[j=0,j<Length[L],j++,NK1=Join[NK1,{Take[L2[[j+1]],i-1]}]];
      L5={};L4={};L3={};L2={};L1={};
      For[j=0,j<Length[L],j++,L1=Join[L1,{Take[L[[j+1]],i]}]];
      For[j=0,j<Length[L],j++,L2=Join[L2,{Transpose[L1[[j+1]]]}]];
      For[j=0,j<Length[L],j++,L3=Join[L3,{Take[L2[[j+1]],i]}]];
      For[j=0,j<Length[L],j++,L4=Join[L4,{Last[L3[[j+1]]]}]];
      For[j=0,j<Length[L],j++,L5=Join[L5,{Most[L4[[j+1]]]}]];
      For[j=0,j<Length[L],j++,lk=Join[lk,{Transpose[{L5[[j+1]]}]}]];
   ];
   Return[{NK1,nkk,lk}]  ];

KSIP[L_List,M_List,N_List,i_]:=KSIP[L,M,N,i]= (* Compute \psi_{i,j+1} (3.21) *)
  Module[{L1={},L2={},L3={},L4={},L5={},iz,gr,j},
   L1=YYP[L,M,N,i-1]; L2=ZZP[L,M,N,i-1]; L3=InvNKP[N,i];
   L4=L3[[3]];gr=Length[L1]+L3[[4]]; iz={};
   L1=DopZero[L1,gr]; L4=DopZero[L4,gr];
   For[j=0,j<gr,j++,
     iz=Join[iz,{Sum[L1[[j-k+1]]L4[[k+1]],{k,0,j}]}];];
   iz=LastZeroP[iz];Return[iz]];

FIP[L_List,M_List,N_List,i_]:=FIP[L,M,N,i]= (* Compute \varphi_{i,j+1} (3.20) *)
 Module[{L1={},L2={},L3={},L4={},L5={},L6={},iz,iz1,Y,Y1,gr,gr1,gr2,gr3,j},
  L1=YYP[L,M,N,i-1]; L2=ZZP[L,M,N,i-1]; L3=InvNKP[N,i]; L4=L3[[1]];
  L6=NKP[N,i][[3]]; L5=TakeFPoly[L,i-1];
  gr1=Max[Length[L1]-1,Length[L2]+Length[L]-1];
  L0=DopZero[FrmPoly[IdentityMatrix[i-1]],gr1];
  L1=DopZero[L1,gr1];Y1={};
  For[j=0,j<gr1,j++,
     Y1=Join[Y1,{Sum[L1[[j-k+1]]L0[[k+1]],{k,0,j}]}]];
  gr2=Length[L2]+Length[L]-1;
  L2=DopZero[L2,gr2];L5=DopZero[L5,gr2];Y = {};
  For[j=0,j<gr2,j++,
     If[Length[Dimensions[L2[[1]]]] == 1,
      Y= Join[Y,{Sum[{L2[[j-k+1]]}.Transpose[L5[[k+1]]],{k,0,j}]}],
      Y= Join[Y,{Sum[L2[[j-k+1]].Transpose[L5[[k+1]]],{k,0,j}]}]
     ]];
  gr3=L3[[2]]+Length[N];L4=DopZero[L4,gr3];
  L6=DopZero[L6,gr3]; iz1={};
  For[j=0,j<gr3,j++,iz1=Join[iz1,{Sum[L4[[j-k+1]].L6[[k+1]],{k,0,j}]}]];
  gr=gr1+gr3; Y=DopZero[Y,gr]; Y1=DopZero[Y1,gr];
  iz1=DopZero[iz1,gr]; iz={};
   For[j=0,j<gr,j++,
       iz=Join[iz,{Sum[(Y1[[j-k+1]]-Y[[j-k+1]]).iz1[[k+1]],{k,0,j}]}]];
  iz=LastZeroP[iz];Return[iz]];

TET[L_List,M_List,N_List,i_]:=TET[L,M,N,i]= (* Compute \Theta_{i,j+1} (3.19) *)
 Module[{L1={},L2={},L3={},L4={},L5={},L6={},iz,iz1,iz2,iz3, gr,gr1,j,k1,k2},
  L1=ZZP[L,M,N,i-1]; L2=InvNKP[N,i];gr1=L2[[4]];
  L2=L2[[3]]; L3=WWP[L,M,N,i]; L4=VVP[L,M,N,i];
  L5=DDP[L,M,N,i];L6=FIP[L,M,N,i]; iz={};
  gr=Length[L1]+gr1+Length[L3]+Length[L4]-1;
  L1=DopZero[L1,gr]; L2=DopZero[L2,gr]; L3=DopZero[L3,gr]; L4=DopZero[L4,gr];
  If[L5=={},For[j=1,j<=i,j++,L5=Append[L5,{{0}}]];];
    L5=DopZero[L5,gr]; iz1={};iz2={};iz3={};
    For[j=0,j<gr,j++,
     iz1=Join[iz1,{Sum[Sum[L1[[j-k-r+1]]L2[[k+1,1]]L3[[r+1]],{k,0,j-r}],{r,0,j}]}];
     If[L5[[1]]*0==={0},
      iz2=Join[iz2,{Sum[Sum[({L5[[j-k-r+1]]}L2[[k+1,1]]).{L4[[r+1]]},{k,0,j-r}],{r,0,j}]}],
      iz2=Join[iz2,{Sum[Sum[(L5[[j-k-r+1]]L2[[k+1,1]]).{L4[[r+1]]},{k,0,j-r}],{r,0,j}]}]]];
   If[L6=={},L6=Table[{0},{k1,1},{k2,i-1}]];
   L6=DopZero[L6,gr];
   For[j=0,j<gr,j++,
       iz3=Join[iz3,{Sum[L6[[j-k+1]].{L4[[k+1]]},{k,0,j}]}]];
   For[j=0,j<gr,j++,
    If[i==2,iz=Join[iz,{{iz1[[j+1]]}-iz2[[j+1]]-iz3[[j+1]]}],
            iz=Join[iz,{iz1[[j+1]]-iz2[[j+1]]-iz3[[j+1]]}]]];
   iz=LastZeroP[iz]; If[iz=={},iz={{0}}];
   Return[iz]];

WPartPoly[L_List,M_List,N_List]:=    (* Implementation of Algorithm 3.1 *)
  Module[{mm,nn,k,rez={},L1={},L2={},L3={},M1={},N1={}},
   {mm,nn}=Dimensions[L];
   A=FrmPoly[L];M1=FrmPoly[M];N1=FrmPoly[N];
   For[k=1,k<=nn-1,k++,
     L1=ZZP[A,M1,N1,k];Print["ZZP=",L1];
     L2=YYP[A,M1,N1,k];
     If[L1===L1*0, rez={L1,{1}}, rez=SimplP[L1,L2]];
     ZZP[A,M1,N1,k]=rez[[1]]; YYP[A,M1,N1,k]=rez[[2]];
     L2=Sum[rez[[2,j]](Variables[L][[1]])^(j-1),{j,1,Length[rez[[2]]]}];
     L1=Sum[rez[[1,j]](Variables[L][[1]])^(j-1),{j,1,Length[rez[[1]]]}];
     L1=VVP[A,M1,N1,k+1]; L2=WWP[A,M1,N1,k+1];
     rez=SimplP[L1,L2]; VVP[A,M1,N1,k+1]=rez[[1]];
     WWP[A,M1,N1,k+1]=rez[[2]];
   ];
   L1=ZZP[A,M1,N1,nn]; L2=YYP[A,M1,N1,nn];  rez=SimplP[L1,L2];
   Print["ZZP[",nn,"]=",rez[[1]]];
   Print["YYP[",nn,"]=",rez[[2]]];
   L2=Sum[rez[[2,j]](Variables[L][[1]])^(j-1),{j,1,Length[rez[[2]]]}];
   L1=Sum[rez[[1,j]](Variables[L][[1]])^(j-1),{j,1,Length[rez[[1]]]}];
   Return[Simplify[L1/L2]//MatrixForm]];

PolLCM[L_List]:=  (* Find the least common multiple *)
  Module[{m=m1=1,j},
    For[j=1,j<=Length[L],j++,
      If[Variables[L]=!={},
         m=PolynomialLCM[m,L[[j]]],
         m=LCM[m,L[[j]]]
    ] ];
    If[Variables[m]=={},
      If[Length[m]=!=0,
         For[j=1,j<=Length[m],j++, m1=LCM[m1,m[[j]]]
      ]  ],
      If[Not[Head[m]=!=List],
         For[j=1,j<=Length[m],j++,   m1=PolynomialLCM[m1,m[[j]]]  ],
         m1=m
    ] ];
    Return[Expand[m1]]  ];

SimplP[M1_List,M2_List]:=
  Module[{p,q,r,vr={},M3=M4={},i},
    p=Sum[M1[[i+1]]*w^i,{i,0,Length[M1]-1}];
    q=Sum[M2[[i+1]]*w^i,{i,0,Length[M2]-1}];
    If[Head[q]=!=List,r=Simplify[p/q],r=Simplify[p/q[[1]]]];
    M3=PolLCM[Denominator[r]];
    If[Variables[M3]=!={}, M4=Expand[Simplify[r*M3]];
      M3=Transpose[FrmPoly[{M3}]][[1]];
      M4=FrmPoly[M4], M4=FrmPoly[r]; M3={1}] ;
    Return[{M4,M3}]];

Delt[L_List,M_List,N_List,i_]:=Delt[L,M,N,i]= (* Compute \Delta_{i,j+1} (3.15),(3.16) *)
  Module[{L1={},L2={},L3={},L4={},L5={},L6={},L7={},gr,gr0,gr1,gr2,
         rez,del1,del2,del21,del22,del23,del24,del25},
 L1=YYP[L,M,N,i-1]; L2=InvNKP[N,i];gr1=L2[[4]];gr2=L2[[2]];
 L2=L2[[3]]; L3=NKP[N,i][[1]]; L4=NKP[N,i][[2]];
 L5=NKP[N,i][[3]]; L6=DDP[L,M,N,i]; L7=FIP[L,M,N,i];
 If[L6=={},L6={Transpose[L5[[1]]]*0}]; If[L7=={},L7={L6[[1]]*0}];
 gr=2*Length[L1]+gr1-1; L1=DopZero[L1,gr]; L2=DopZero[L2,gr];del1={};
 For[j=0,j<gr,j++, del1=Join[del1,{Sum[Sum[L1[[j-k-r+1]]L1[[k+1]]L2[[r+1]],
                                                  {k,0,j-r}],{r,0,j}]}];];
 gr0=3*Length[L1]+gr2+gr;L1=DopZero[L1,gr0];
 L2=DopZero[L2,gr0]; L3=DopZero[L3,gr0];L4=DopZero[L4,gr0];
 L5=DopZero[L5,gr0]; L6=DopZero[L6,gr0];L7=DopZero[L7,gr0];
 del2={};del21={};del22={};del23={};del24={};del25={};
 If[i==2,For[j=0,j<gr0,j++,
     del21=Join[del21,{Sum[Sum[Sum[L4[[j-k-r-t+1]]L1[[k+1]]L1[[r+1]]L2[[t+1]],
                            {k,0,j-t-r}],{r,0,j-t}],{t,0,j}]}];
     del22=Join[del22,Sum[Sum[Sum[Transpose[{L6[[j-k-r-t+1]]}].
                 L3[[k+1]][[1]]L6[[r+1]]L2[[t+1]],{k,0,j-t-r}],{r,0,j-t}],{t,0,j}]];
     del23=Join[del23,Sum[Sum[Sum[Transpose[{L6[[j-k-r-t+1]]}].
                 L5[[k+1]][[1]]L1[[r+1]]L2[[t+1]],{k,0,j-t-r}],{r,0,j-t}],{t,0,j}]];
     del24=Join[del24,Sum[Sum[Sum[Transpose[L5[[j-k-r-t+1]]].
                 L6[[k+1]]L1[[r+1]]L2[[t+1]],{k,0,j-t-r}],{r,0,j-t}],{t,0,j}]];
     del25=Join[del25,Sum[Sum[Transpose[L5[[j-k-r+1]]].
                 L7[[k+1]]L1[[r+1]],{k,0,j-r}],{r,0,j}]]],
   For[j=0,j<gr0,j++,
    del21=Join[del21,{Sum[Sum[Sum[L4[[j-k-r-t+1]]L1[[k+1]]L1[[r+1]]L2[[t+1]],
                      {k,0,j-t-r}],{r,0,j-t}],{t,0,j}]}];
    del22=Join[del22,Sum[Sum[Sum[Transpose[L6[[j-k-r-t+1]]].L3[[k+1]].
             L6[[r+1]]L2[[t+1]],{k,0,j-t-r}],{r,0,j-t}],{t,0,j}]];
    del23=Join[del23,Sum[Sum[Sum[Transpose[L6[[j-k-r-t+1]]].
             L5[[k+1]]L1[[r+1]]L2[[t+1]],{k,0,j-t-r}],{r,0,j-t}],{t,0,j}]];
    del24=Join[del24,Sum[Sum[Sum[Transpose[L5[[j-k-r-t+1]]].
             L6[[k+1]]L1[[r+1]]L2[[t+1]],{k,0,j-t-r}],{r,0,j-t}],{t,0,j}]];
    del25=Join[del25,Sum[Sum[Transpose[L5[[j-k-r+1]]].
             L7[[k+1]]L1[[r+1]],{k,0,j-r}],{r,0,j}]]]];
 del2=del21+del22-del23-del24-del25; del1=LastZeroP[del1];del2=LastZeroP[del2];
 rez=SimplP[del1,del2];
 Return[rez]];

(* ------------ Compute inverse of N --------------*)

NNinv[N_List,i_]:=NNinv[N,i]=  (*   Compute (3.33),(3.34)  *)
  Module[{L0={},L1={},L2={},L3={},L4={},L5={},Y={},e={},f={},g={},rez},
   L1=NKP[N,i]; L2=L1[[2]];
   If[i==1, Print["Ninv[",i,"]=",{{{1}},L2}];Return[{{{1}},L2}],
     Y={}; L3=EEI[N,i][[2]]; L4=FFI[N,i][[2]]; L5=GII[N,i][[2]];
     L6=EEI[N,i][[1]]; L7=FFI[N,i][[1]]; L8=GII[N,i][[1]];
     gr=Length[L3]+Length[L4]+Length[L5];
     L3=DopZero[L3,gr]; L4=DopZero[L4,gr]; L5=DopZero[L5,gr];
     For[j=0,j<gr,j++,
       Y=Join[Y,Sum[Sum[{L3[[j-k-r+1]]}L4[[k+1,1]]L5[[r+1,1]],{k,0,j-r}],{r,0,j}]]];
     Y=LastZeroP[Y]; e={};
     gr=Length[L6]+Length[L4]+Length[L5]; L6=DopZero[L6,gr];
     L4=DopZero[L4,gr]; L5=DopZero[L5,gr];
     For[j=0,j<gr,j++,
      e=Join[e,{Sum[Sum[L6[[j-k-r+1]]L4[[k+1,1]]L5[[r+1,1]],{k,0,j-r}],{r,0,j}]}]];
     If[LastZeroP[e]==={},e={e[[1]]},e=LastZeroP[e]];
     f={}; gr=Length[L7]+Length[L3]+Length[L5]; L7=DopZero[L7,gr];
     L3=DopZero[L3,gr]; L5=DopZero[L5,gr];
     For[j=0,j<gr,j++,
        f=Join[f,{Sum[Sum[L7[[j-k-r+1]]L3[[k+1]]L5[[r+1,1]],{k,0,j-r}],{r,0,j}]}]];
     If[LastZeroP[f]==={},f={f[[1]]},f=LastZeroP[f]];
     f=DopZero[f,Length[e]]; g={}; gr=Length[L8]+Length[L3]+Length[L4];
     L8=DopZero[L8,gr];L3=DopZero[L3,gr]; L5=DopZero[L5,gr];
     For[j=0,j<gr,j++,
          g=Join[g,{Sum[Sum[L8[[j-k-r+1]]L3[[k+1]]L5[[r+1,1]],{k,0,j-r}],{r,0,j}]}]];
     If[LastZeroP[g]==={},g={g[[1]]},g=LastZeroP[g]];
     iz=FrmPoly[FormE[e,f,g]]; rez=SimplP[iz,Y];
     Return[rez]]];

GII[N_List,i_]:=GII[N,I]=   (*   Compute (3.27),(3.28)  *)
  Module[{L0={},L1={},L2={},L3={},L4={},L5={},iz,iz1,iz2,gr,j,k,r,rez},
   L0=NNinv[N,i-1];L1=NKP[N,i];L2=L1[[2]];L3=L1[[3]];
   L4=L0[[1]];L5=L0[[2]];gr=Length[L4]+Length[L5]+2*Length[N];
   L2=DopZero[L2,gr];L4=DopZero[L4,gr];L5=DopZero[L5,gr];
   L3=DopZero[L3,gr];iz1={};iz2={};
   For[j=0,j<gr,j++,
     iz2=Join[iz2,{Sum[L2[[j-k+1]]L5[[k+1]],{k,0,j}]}]];
   For[j=0,j<gr,j++,
     If[i<=2,iz1=Join[iz1,{Sum[Sum[-(Transpose[L3[[j-k-r+1]]].L4[[k+1]])
                                       .L3[[r+1]],{k,0,j-r}],{r,0,j}]}],
       iz1=Join[iz1,Sum[Sum[-Transpose[L3[[j-k-r+1]]].L4[[k+1]]
                                       .L3[[r+1]],{k,0,j-r}],{r,0,j}]];]];
   iz=iz2+iz1;
   Return[{LastZeroP[L5],LastZeroP[iz]}]];

FFI[N_List,i_]:=FFI[N,i]=    (*   Compute (3.29),(3.30)  *)
  Module[{L0={},L1={},L2={},L3={},L4={},L5={},iz,iz1,gr,j,k,r,rez},
   L0=NNinv[N,i-1]; L1=NKP[N,i]; L2=GII[N,i][[2]];
   L3=L1[[3]]; L4=L0[[1]]; L5=L0[[2]]; gr=Length[L4]+Length[N];
   iz={}; L3=DopZero[L3,gr]; L4=DopZero[L4,gr];
   For[j=0,j<gr,j++, iz=Join[iz,{Sum[L4[[j-k+1]].L3[[k+1]],{k,0,j}]}];];
   If[LastZeroP[iz]==={},iz={iz[[1]]},iz=LastZeroP[iz]];
   Return[{-iz,LastZeroP[L2]}]];

EEI[N_List,i_]:=EEI[N,i]=   (*   Compute (3.31),(3.32)  *)
  Module[{L0={},L1={},L2={},L3={},L4={},L5={},L6={},L7={},L8={},
        s1={},s2={},iz,iz1,gr,j,k,r,rez},
   L0=NNinv[N,i-1]; L1=L0[[1]]; L2=L0[[2]];
   L5=FFI[N,i][[1]]; L6=FFI[N,i][[2]]; L7=GII[N,i][[1]]; L8=GII[N,i][[2]];
   gr=Max[Length[L1]+Length[L7]+2*Length[L8],Length[L2]+Length[L8]+2*Length[L7]];
   iz={}; L1=DopZero[L1,gr]; L7=DopZero[L7,gr]; L6=DopZero[L6,gr];
   L5=DopZero[L5,gr]; L2=DopZero[L2,gr]; s1={};
   For[j=0,j<gr,j++,
     If[Length[Dimensions[L5[[1]]]]==1,
       s1=Join[s1,{{Sum[Sum[L1[[j-k-r+1]]L7[[k+1]]L6[[r+1]],{k,0,j-r}],{r,0,j}]}}],
       s1=Join[s1,{Sum[Sum[L1[[j-k-r+1]]L7[[k+1]]L6[[r+1,1]],{k,0,j-r}],{r,0,j}]}]]];
   s2={};
   For[j=0,j<gr,j++,
     If[Length[Dimensions[L5[[1]]]]==1,
       s2=Join[s2,{{Sum[Sum[L2[[j-k-r+1]]L5[[k+1]]L5[[r+1]],{k,0,j-r}],{r,0,j}]}}],
       s2=Join[s2,{Sum[Sum[L2[[j-k-r+1]]L5[[k+1]].Transpose[L5[[r+1]]],{k,0,j-r}],{r,0,j}]}]]];
   iz=s1+s2; gr1=Length[L7]+Length[L2]+Length[L6]; iz1={};
   L2=DopZero[L2,gr1]; L7=DopZero[L7,gr1]; L6=DopZero[L6,gr1];
   For[j=0,j<gr1,j++,
         iz1=Join[iz1,{Sum[Sum[L2[[j-k-r+1]]L7[[k+1]]L6[[r+1,1]],{k,0,j-r}],{r,0,j}]}]];
   Return[{LastZeroP[iz],LastZeroP[iz1]}]];

FormE[e_List,f_List,g_List]:=
   Module[{e1,f1,g1,Y,Y1,Y2,i},
   e1=Sum[e[[i+1]]*w^i,{i,0,Length[e]-1}];
   f1=Sum[f[[i+1]]*w^i,{i,0,Length[f]-1}];
   g1=Sum[g[[i+1]]*w^i,{i,0,Length[g]-1}];
   If[Head[g1]=!=List,g1={g1}];
     Y1=0*e1;
     For[j=1,j<=Length[e1],j++,
       If[Length[f1]==1,Y1[[j]]=Append[e1[[j]],f1[[j]]],
          Y1[[j]]=Join[e1[[j]],f1[[j]]]]];
      If[Length[f1]==1,Y2={Join[f1,g1]},
          Y2={Join[Transpose[f1][[1]],g1]}];
      Y=Join[Y1,Y2];
      Return[Y]];

InvNKP[L_List,i_]:=InvNKP[L,i]=Module[{L0={},rez,iz},  (* Compute (3.35)  *)
   L0=NNinv[L,i-1]; rez=SimplP[L0[[1]],L0[[2]]];
   If[i==2,iz={{rez[[1]]},Length[rez[[1]]]-1,Transpose[{rez[[2]]}],Length[rez[[2]]]-1},
    iz={rez[[1]],Length[rez[[1]]]-1,Transpose[{rez[[2]]}],Length[rez[[2]]]-1}];
Return[iz]];
\end{verbatim}\end{scriptsize}

\enddocument
\bye